\documentclass[12pt]{article}

\usepackage[textsize=tiny]{todonotes}

\usepackage{dcolumn}
\usepackage{amsbsy}
\usepackage{amsmath}
\usepackage{amsthm}

\usepackage{amssymb}
\usepackage{braket}
\usepackage{color}
\usepackage{fullpage}
\usepackage{graphicx} 
\usepackage{caption}
\usepackage{bm}

\usepackage{cleveref}
\usepackage{subfigure}

\usepackage{comment}
\newtheorem{rem}{Remark}

\newcommand{\bx}{\bm{x}}

\newcommand{\bD}{\mathbf{D}}

\newcommand{\bH}{\bm{H}}
\newcommand{\bI}{\bm{I}}
\newcommand{\bK}{\bm{K}}

\newcommand{\bP}{\mathbf{P}}
\newcommand{\bQ}{\mathbf{Q}}
\newcommand{\bR}{\bm{R}}

\newcommand{\bX}{\mathbf{X}}

\begin{document}
\title{A study of deformation localization in nonlinear elastic lattices}
\author{ Raj Kumar Pal$^{a}$, Federico Bonetto$^{b}$, Luca Dieci$^{b}$ and Massimo Ruzzene$^{a,c,*}$ \\
{\small $^a$ School of Aerospace Engineering, Georgia Institute of Technology, Atlanta GA 30332, USA}\\
{\small $^b$ School of Mathematics, Georgia Institute of Technology, Atlanta GA 30332, USA}\\
{\small $^c$ School of Mechanical Engineering, Georgia Institute of Technology, Atlanta GA 30332, USA}\\
{\small$^*$Corresponding author. E-mail: ruzzene@gatech.edu}}


\maketitle

\begin{abstract}
The paper investigates localized deformation patterns resulting from the onset 
of instabilities in lattice structures. The study is motivated by previous 
observations on discrete hexagonal lattices, where the onset of non-uniform, 
quasi-static deformation patterns was associated with the loss of convexity of 
the interaction potential, and where a variety of  localized deformations were 
found depending on loading configuration, lattice parameters and boundary 
conditions. These observations are here conducted on other lattice structures, 
with the goal of identifying models of reduced complexity that are able to 
provide insight into the key parameters that govern the onset of 
instability-induced localization. To this end, we first consider a 
two-dimensional square lattice consisting of point masses connected by in-plane 
axial springs and vertical ground springs. Results illustrate that depending on 
the choice of spring constants and their relative values, the lattice exhibits 
in-plane or out-of plane instabilities leading to folding and unfolding. This 
model is further simplified by considering the one-dimensional case of a 
spring-mass chain sitting on an elastic foundation, which may be considered as a 
discretized description of an elastic beam supported by an elastic substrate. A 
bifurcation analysis of this lattice identifies the stable and unstable branches 
and illustrates its hysteretic and loading path-dependent behaviors. Finally, 
the lattice is further reduced to a minimal four mass model which 
undergoes a folding/unfolding process qualitatively similar to the same 
process in the central part of a longer chain, helping our understanding 
of localization in more complex systems. In contrast to 
the widespread assumption that localization is induced by defects or 
imperfections in a structure, this work illustrates that such phenomena can 
arise in perfect lattices as a consequence of the mode-shapes at the bifurcation 
points. 
\end{abstract}

\clearpage
\section{Introduction} 

Localized deformations resulting form instabilities arise naturally in a wide 
range of physical systems, and across multiple length scales. Manifestations 
include plastic twinning in metals~\cite{yoo1981slip}, localized buckling of 
epithelial cells in biological media~\cite{murisic2015discrete}, and Chevron 
folds in rocks~\cite{ramsay1974development}, among others. The study of 
instabilities, both at the microstructural scale in materials and at the 
macroscopic structural level, is an area of renewed interest, and a timely 
research topic. Several studies have focused on exploiting the formation of 
patterns resulting from instabilities in engineered materials and structures to 
enable a variety of functionalities that are useful to applications ranging from 
flexible electronics to architected adaptive 
materials~\cite{kochmann2017exploiting}. Instabilities and the ensuing pattern 
formations, or topological changes, for example, govern phase transitions in 
materials such as shape memory alloys~\cite{bhattacharya2003microstructure} and  
attempts have been made at replicating similar principles in structural 
components. Extensive research has also been devoted to the study of 
instabilities in thin films on soft 
substrates~\cite{Xu:2015aa,Pocivavsek:2008aa}, periodic 
composites~\cite{geymonat1993homogenization} and 
lattices~\cite{Kane:2014aa,Anderson:1984aa,Paulose:2015aa}. Relevant examples 
include the investigation of global pattern formation in periodic composites and 
lattices~\cite{triantafyllidis1998onset}, and the onset of herringbone patterns 
in compressed thin films ~\cite{hutchinson2004herringbone}. More recently, 
Bertoldi and Boyce demonstrated how instabilities in soft polymers induce 
changes in the frequency band structure of periodic phononic 
systems~\cite{bertoldi2008mechanically}, while Pal et 
al.~\cite{pal2016continuum,pal2016effect} investigated the static and dynamic 
properties of hexagonal lattices and demonstrated that instabilities can lead to 
the surface confinement of elastic waves. Engineered defect distributions that 
induce desired deformation patterns in thin shells have been investigated 
in~\cite{lee2016geometric}, where the onset of instabilities is shown to be 
associated with the breaking of discrete lattice translational symmetry and can 
be predicted by a phonon stability analysis on a unit cell. 

Of particular interest to the present study is the investigation of localized 
deformations associated with the onset of instabilities. Methodologies for the 
prediction and design of localized patterns are the objectives of numerous 
studies and are considered open challenges towards the understanding of failure 
as well as the engineering of desired interfaces that act as tunable elastic 
waveguides. The investigation of localization resulting from post-buckling in 
lattices and periodic media is presented for example 
in~\cite{pal2016continuum,combescure2016post}.  Although the general principles 
leading to localization in continuous media, which manifests as discontinuous 
strain distributions, is typically assessed by examining the loss of ellipticity 
in the Hessian of the strain energy~\cite{rudnicki1975conditions}, its relation 
to the microstructure and the effect of the macroscopic geometry including 
boundary conditions remain elusive. Indeed, in contrast to a phonon stability 
analysis on a single unit cell for identifying global instabilities, no such 
recipe exists for localization.  Several 
studies~\cite{gibson1999cellular,papka1994plane} have demonstrated, both 
numerically and experimentally, how buckling at the microstructural level 
evolves into localized deformations. In this context, notable are the works of 
Papka and Kyriakides~\cite{papka1998experiments,papka1999biaxial} who 
investigated the crushing of honeycomb cellular lattices under a variety of 
loading conditions. More recently, d'Avila et al.~\cite{d2016localization} have 
demonstrated that the onset of localization in a periodic composite depends on 
the effective tangent stiffness of the composite and occurs only if this 
stiffness in the loading direction is negative. 

\begin{figure}
\centering
\subfigure[]{\includegraphics[width=0.45\textwidth]{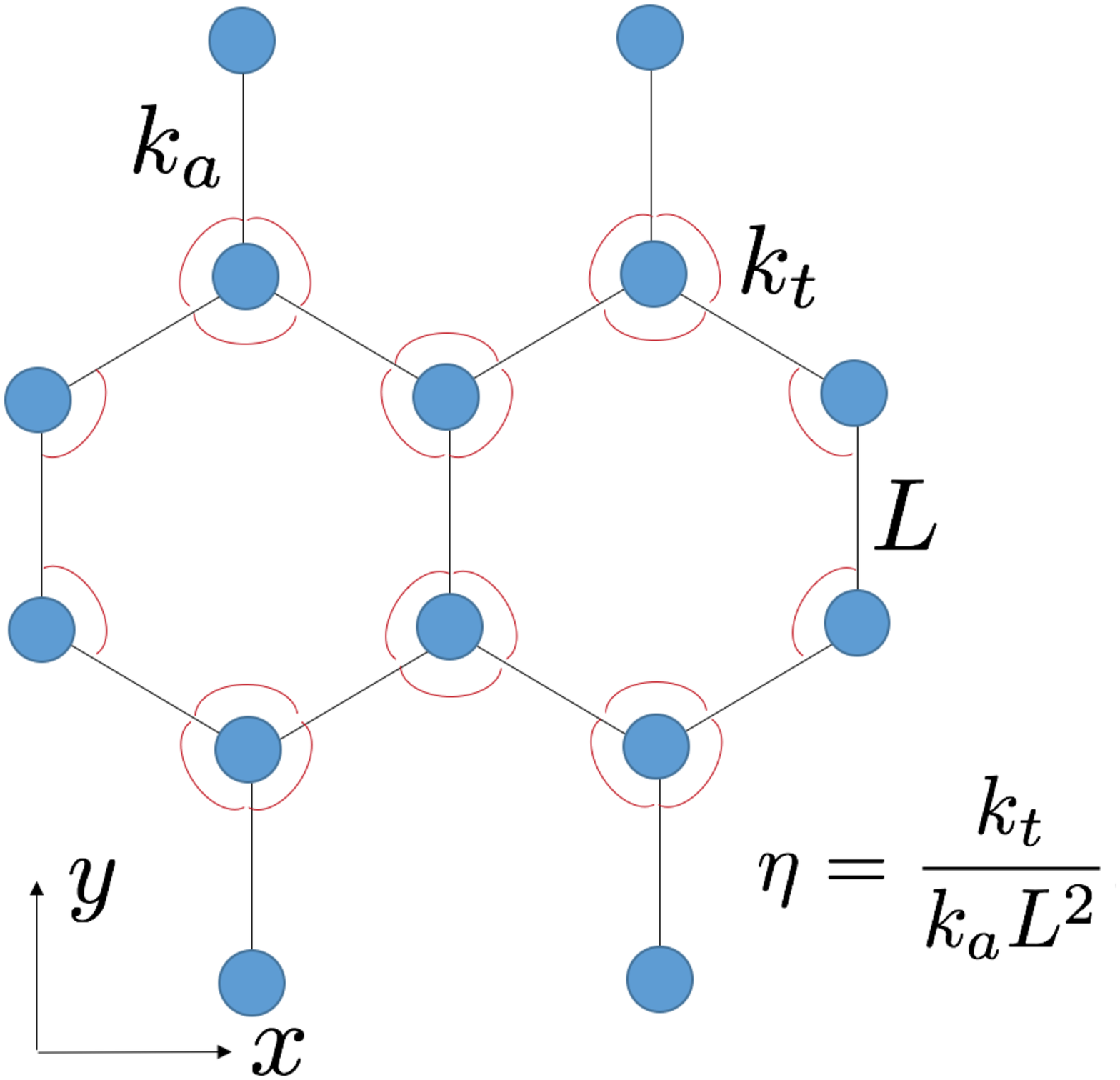} \label{fig1a}}
\subfigure[]{\includegraphics[width=0.6\textwidth]{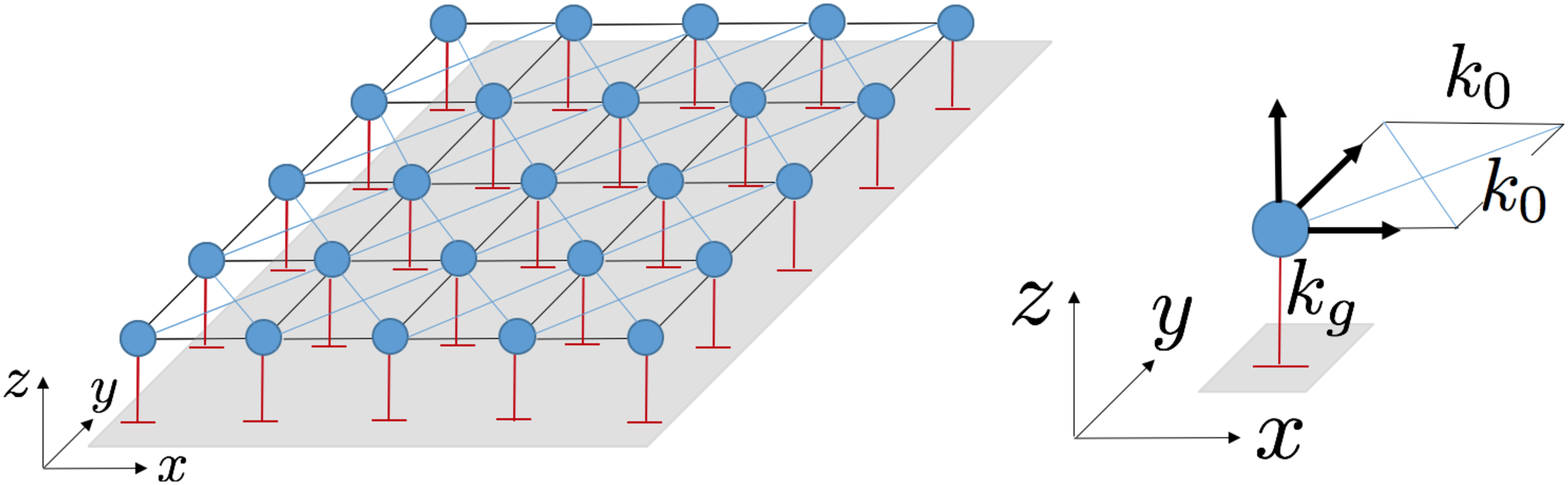} \label{fig1b}}\\
\subfigure[]{\includegraphics[width=0.45\textwidth]{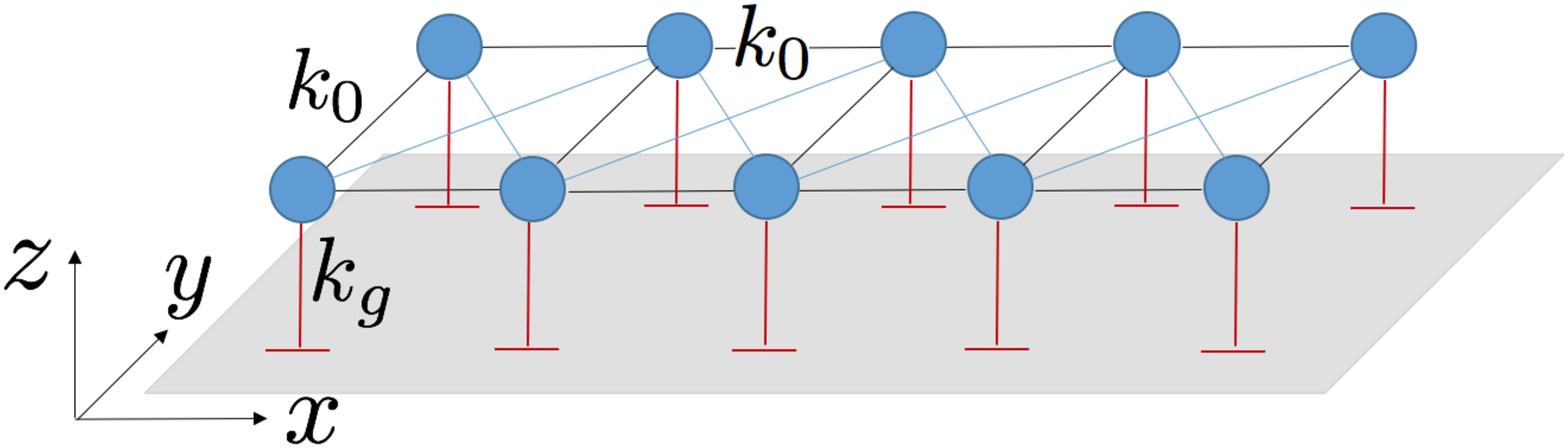} \label{fig1c}}
\subfigure[]{\includegraphics[width=0.45\textwidth]{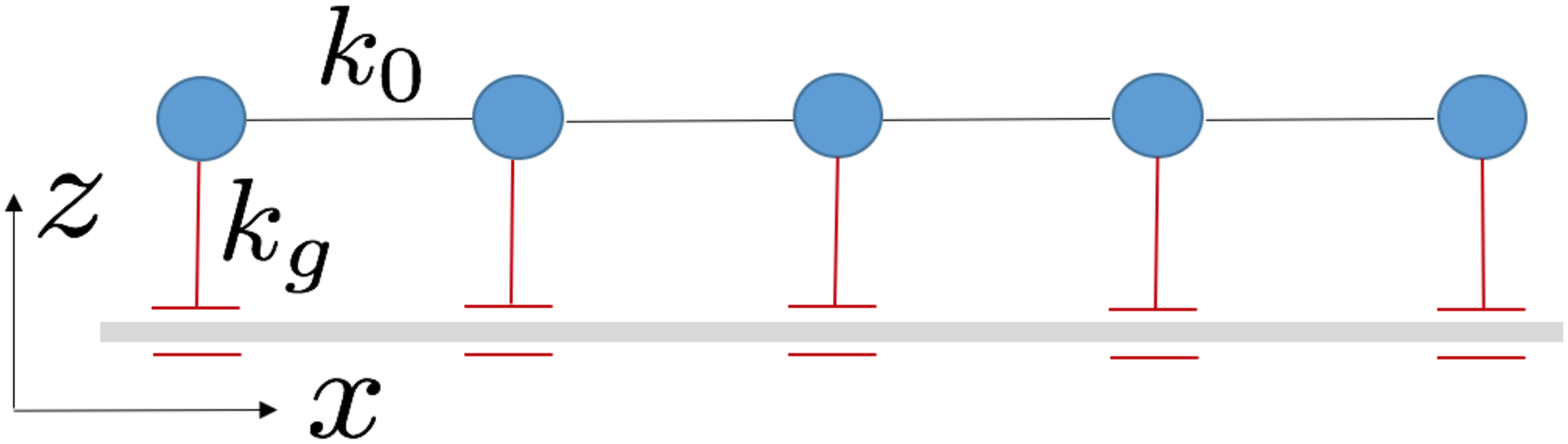} \label{fig1d}}
\caption{Summary of discrete lattices considered for observation and analysis of 
instability-induced localization: hexagonal lattice investigated 
in~\cite{pal2016continuum} (a), square lattice with ground springs (left) and 
detail of unit cell (right) (b), one-dimensional square lattice strip (c), and 
1D spring mass chain (d).} \label{fig1}
\end{figure}

The current study is motivated by the observation of localized deformation 
patterns following the onset of instabilities in discrete, hexagonal 
lattices~\cite{pal2016continuum}. Such deformations occur as a result of the 
presence of nonlinearities corresponding to large displacements, but are not 
associated with the existence of defects and imperfections, which differs from 
the typical assumptions made in most studies on localization. Some of the 
localized deformation patterns observed in~\cite{pal2016continuum} are here 
presented as background to the investigation presented herein. Specifically, the 
paper focuses on progressively simpler lattices with the objective of 
identifying configurations that are characterized by localized deformations and 
that possibly lend themselves to analytical treatment and may provide useful 
insight into the most relevant parameters that govern the behavior of interest. 
An overview of the considered lattice configurations is shown in 
Fig.~\ref{fig1}, which presents the original in-plane hexagonal lattice 
(Fig.~\ref{fig1a}) and an elastically supported square lattice capable of both 
in-plane and out-of-plane motion (Fig.~\ref{fig1b}). The dimensionality of the 
latter problem is further reduced by extracting the single lattice strip shown 
in Fig.~\ref{fig1c}, which then leads to the elementary case of the elastically 
supported one-dimensional (1D) spring mass chain shown in Fig.~\ref{fig1d}. The 
hexagonal lattice consists of masses connected by longitudinal springs denoted 
as connecting lines in the Fig.~\ref{fig1a}, and includes angular springs, 
denoted by the red arcs, that provide a restoring moment proportional to the 
relative rotations of neighboring springs. In all other figures in 
Fig.~\ref{fig1}, the thin lines describe longitudinal springs, that provide a 
restoring force aligned with the spring itself and proportional to the relative 
displacements of the two nodes it connects. The onset of localized deformation 
is initially based on observation of the deformed equilibrium configurations 
resulting from the application of strains, which are evaluated through a 
Newton-Raphson procedure. For the low dimensional cases of 
Fig.~\ref{fig1c},~\ref{fig1d}, a linear bifurcation analysis and the application 
of a shooting method support the interpretation of the observed solutions, the 
prediction of stability characteristics, the evolution of the equilibrium 
deformed configurations for increasing levels of applied strain as well as the 
evaluation of loading path-dependent or hysteresis behaviors. A rich set of 
equilibrium solutions which evolve from global patterns, to strongly localized 
ones that are associated with folding and unfolding characteristics. These are 
reminiscent of experimental observations made on thin elastic membranes on soft 
supports as presented for example in~\cite{Pocivavsek:2008aa}. This study 
illustrates how deformed localization occurs as a result of nonlinearities 
associated with large deformations. It also shows that a systems as 
simple as a 4 mass 1D chain can help understand this phenomenon. This 
suggests 
that simple configurations as investigated herein may provide the basis for the 
formulations of a general framework for the investigation of 
instability-induced localization, along with guidelines for the design of 
assemblies with engineered localization patterns. 

The paper is organized as follows. Following this introduction, 
Sec.~\ref{hexNumerSec} provides an overview of localization patterns observed in 
the hexagonal lattice of Fig.~\ref{fig1a}. Next, Sec.~\ref{SquareSec} presents 
two distinct kinds of instability that arise in square lattices interacting with 
ground springs: a localized deformation and a globally uniform pattern. 
Section~\ref{chainSec} investigates this localized deformation in an analogous 
one-dimensional lattice using a combination of numerical simulations and 
bifurcation analysis. Finally, the conclusions of this work are summarized in 
Sec.~\ref{concSec}.

\section{Background: instability-induced localization in hexagonal lattices}\label{hexNumerSec}

Instabilities in lattices leading to localized deformations or global pattern 
formation depend on the interplay of several factors, including lattice 
geometry, material parameters, boundary and loading conditions. The effect of 
some of these factors is here briefly illustrated as background and motivation 
for the investigations to follow. We consider the quasi-static behavior of the 
hexagonal lattice of Fig.~\ref{fig1a}, which consists of point masses connected 
by linear longitudinal springs $k_a$, and includes angular springs $k_t$ that 
oppose the change in angle between neighboring 
springs~\cite{pal2016continuum,pal2016effect}. The lattice is constrained to 
deform in the $x,y$ plane and undergoes large displacements resulting from the 
imposed set of displacements. We determine the equilibrium configuration of a 
finite assembly of $32\times 23$ unit cells by minimizing the potential energy 
of the lattice at each incremental step of the prescribed displacement using a 
Newton-Raphson procedure. This procedure is described in detail 
in~\cite{pal2016continuum}. The behavior of the lattice is controlled by a 
single nondimensional parameter $\eta=k_t/k_a L^2$, which quantifies the 
relative values of the axial and torsional springs, with $L$ being the distance 
between the two sub-lattice sites in a unit cell. The case of $\eta=6 \times 
10^{-3}$ here considered corresponds to a lattice characterized by a non-convex 
energy and that exhibits complex deformation patterns under uniform loading 
conditions due to instabilities. 

\begin{figure}
\centering
\subfigure[]{\includegraphics[width=0.44\textwidth]{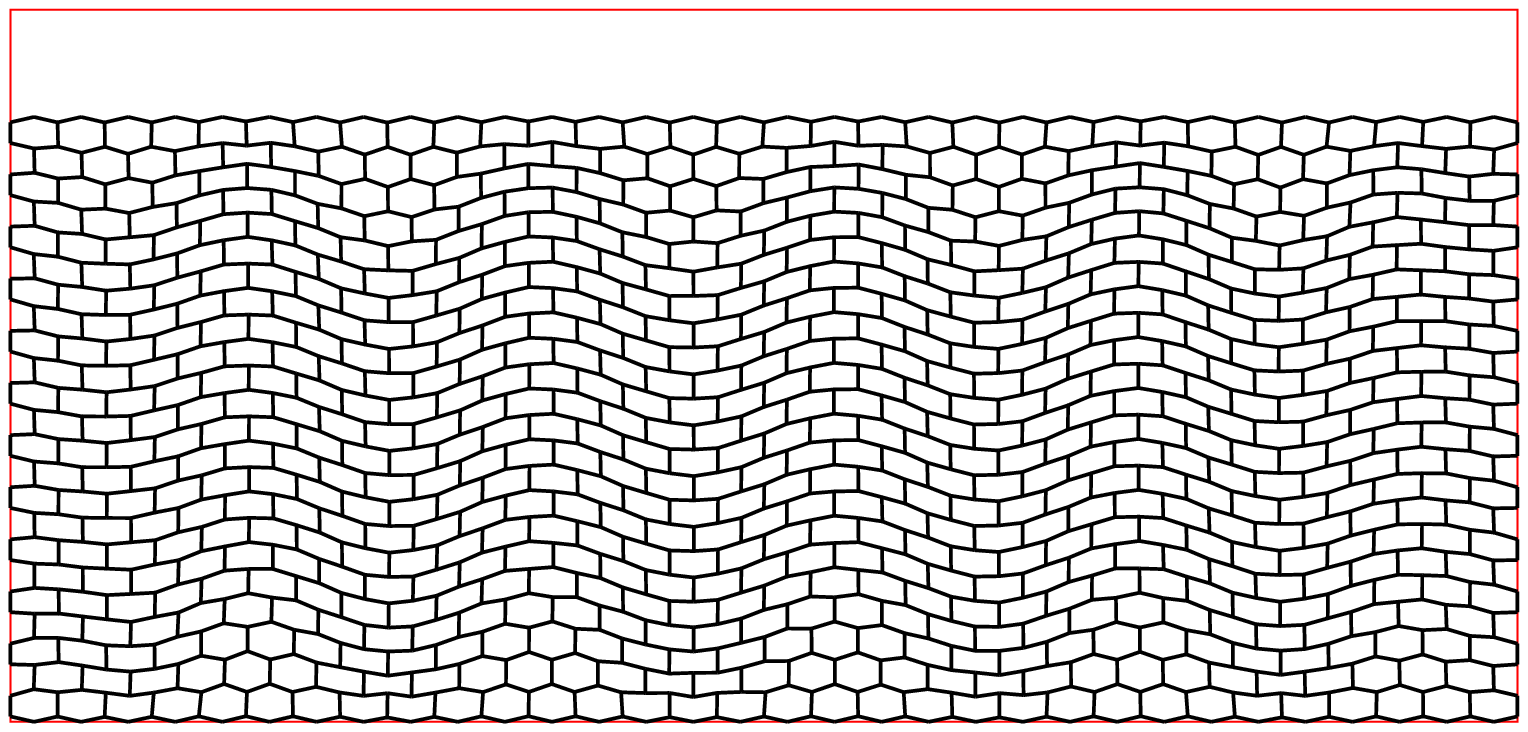} \label{fig2a}}
\subfigure[]{\includegraphics[width=0.463\textwidth]{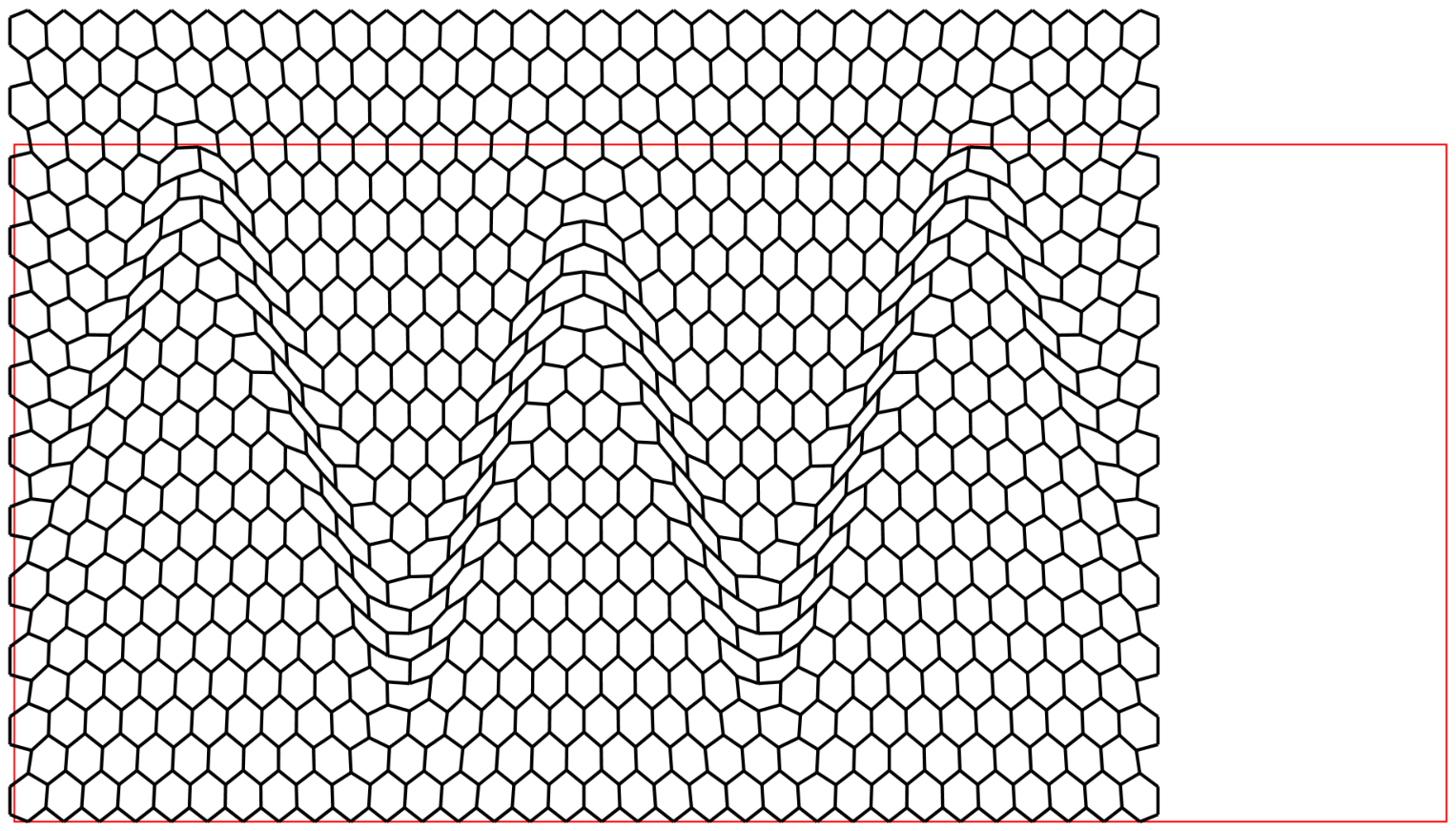}
\label{fig2b}}\\
\subfigure[]{\includegraphics[width=0.463\textwidth]{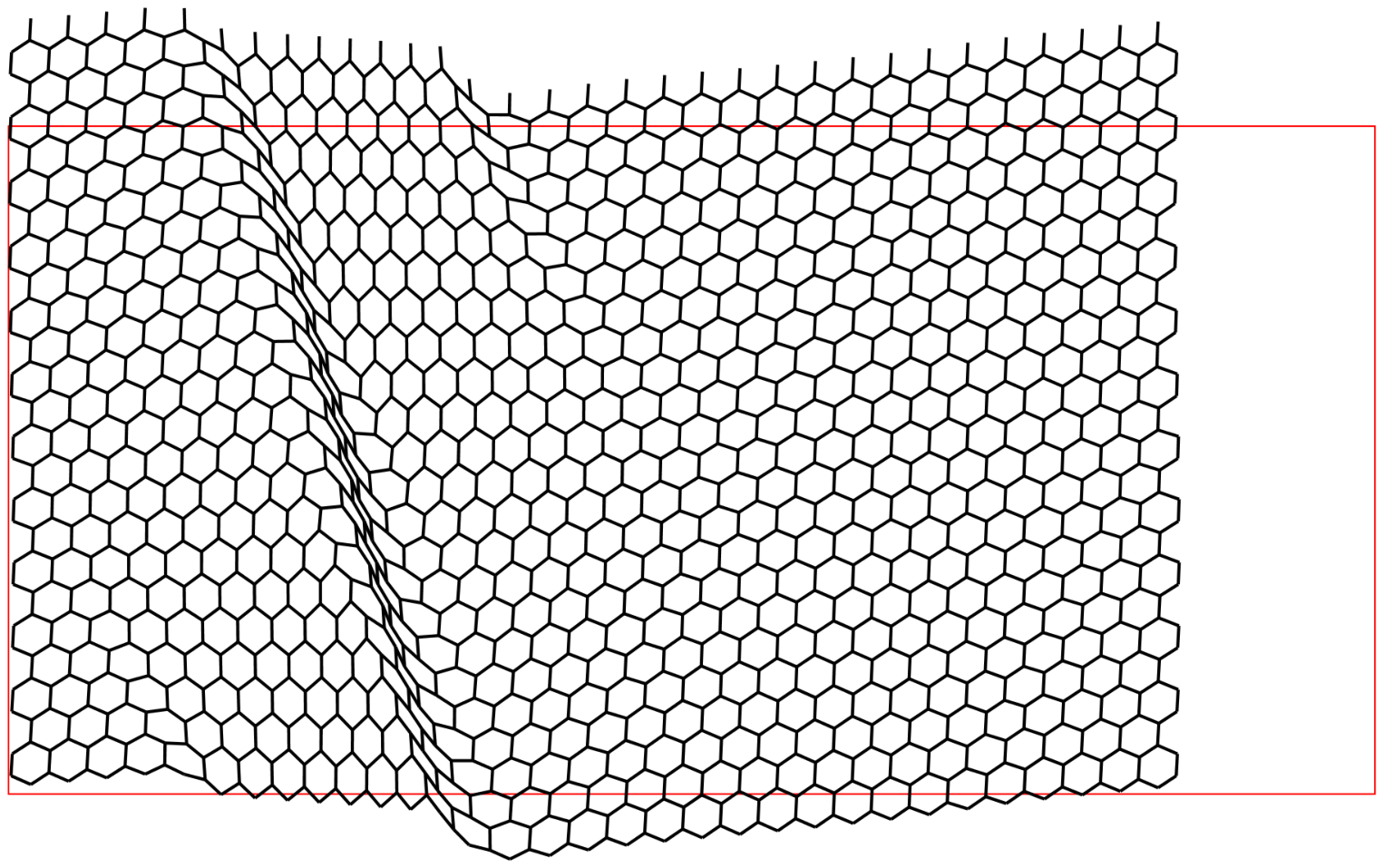}
\label{fig2c}}
\caption{Deformed configurations having different localization patterns for 
different boundary conditions. (a) Biaxial loading with displacement prescribed 
on all boundary nodes; (b) compression prescribed on horizontal direction and 
periodic conditions along vertical boundary nodes; (c) compression prescribed on 
horizontal direction, free to expand and periodic conditions in vertical 
direction.} 
\label{fig2}
\end{figure}

We first illustrate the lattice deformation corresponding to an imposed vertical 
displacement leading to a total normal strain for the finite lattice of 
$\delta_{y} = -0.15$. As the imposed displacement is progressively increased, 
the deformation evolves from being initially affine (linear in $x$ and $y$) to 
subsequently achieving global patterns as in Fig.~\ref{fig2a}. These patterns, 
associated with a long wavelength instability, break the translation invariance 
of the displacement field in the $x$-direction and their onset coincides with 
with the loss of rank-one convexity in the potential energy of a single unit 
cell~\cite{pal2016continuum}. Next, we discuss the case of a bi-axial loading 
corresponding to imposed displacements along both the horizontal and vertical 
direction, leading to normal strains for the lattice domain respectively equal 
to $\delta_{x} = -0.2$ and $\delta_{y} = 0.2$. The resulting deformed 
configuration shown in Fig.~\ref{fig2b} is characterized by a localized 
deformation along a ``zig-zag" interface. This pattern of localized deformation 
is attributed to the competition between two factors: localized deformation 
along the unit vector direction and the imposed affine boundary conditions.

Finally, we illustrate how the localization pattern can vary with boundary 
conditions. Figure~\ref{fig2c} displays the deformation pattern when periodic 
boundary condition\footnote{The periodic constraint relating a top node $\bx_t$ 
with its corresponding bottom node $\bx_b$ takes the form $\bx_t - \bx_b = 
(0,w)$. The width $w$ is constant across the lattice, but it is not constrained 
and allowed to change from its initial value. As a result, the lattice expands 
in the $y$ direction due to Poisson effect} are imposed along with a prescribed 
horizontal normal strain $\delta_{x}=-0.12$. In contrast to the case in 
Fig.~\ref{fig2b}, localization occurs along  one of the lattice vector 
directions, at an angle $2\pi/3$ from the $x$-axis. Indeed the change in 
localization pattern compared to the previous case is attributed to relaxing the 
affine displacement constraint on the top and bottom surface nodes. Thus the 
deformation patterns that result in these nonlinear lattices depend on their 
geometry, lattice parameters or material properties, loading and boundary 
conditions. Elucidating the interplay between these factors is a daunting 
challenge, and is the subject of ongoing investigations. To address some of the 
questions that the observations presented above pose, we here focus on the 
effects of lattice parameters under a single, uni-axial loading condition. To 
this end, we elect to investigate the lattices in Fig.~\ref{fig1b}-\ref{fig1d}, 
which are characterized by a simpler geometry described by orthogonal lattice 
vectors, but which however have a potentially richer behavior related to the 
ability of deform both in-plane and out-of-plane. The interaction of these 
deformations and their relation to the onset of various instabilities will be 
investigated in lattices of increasing simplicity in an attempt to obtain 
general insight into the unique mechanism observed herein.

\section{Square lattice model}\label{SquareSec}
Let us consider a square lattice of $N\times M$ nodes, that is $(N-1)\times 
(M-1)$ unit cells, which includes point masses at the nodes and three types of 
springs connecting them. In contrast to the previous hexagonal lattice case 
where the masses were restricted to move in the $xy$-plane, here each point mass 
has $3$ translational degrees of freedom and can move in three spatial 
dimensions. The diagonal springs and the springs connecting nearest neighbors 
both have stiffness $k_0$. There is a ground spring of stiffness $k_g$, which 
resists motion in the out-of-plane $z$ direction. The quasi-static behavior of a 
lattice of a given dimension is governed by a single nondimensional stiffness 
parameter, $\gamma = k_g/k_0$. In all the subsequent calculations in this work, 
we investigate the behavior of a lattice under uniaxial compression. Both the 
in-plane displacement components are prescribed at all the boundary nodes 
corresponding to this uniaxial strain, while the out-of-plane displacement 
component is free at all the nodes. 

\subsection{In-plane and out-of-plane instability}

\begin{figure}
\centering
\subfigure[]{
\includegraphics[width=0.4\textwidth]{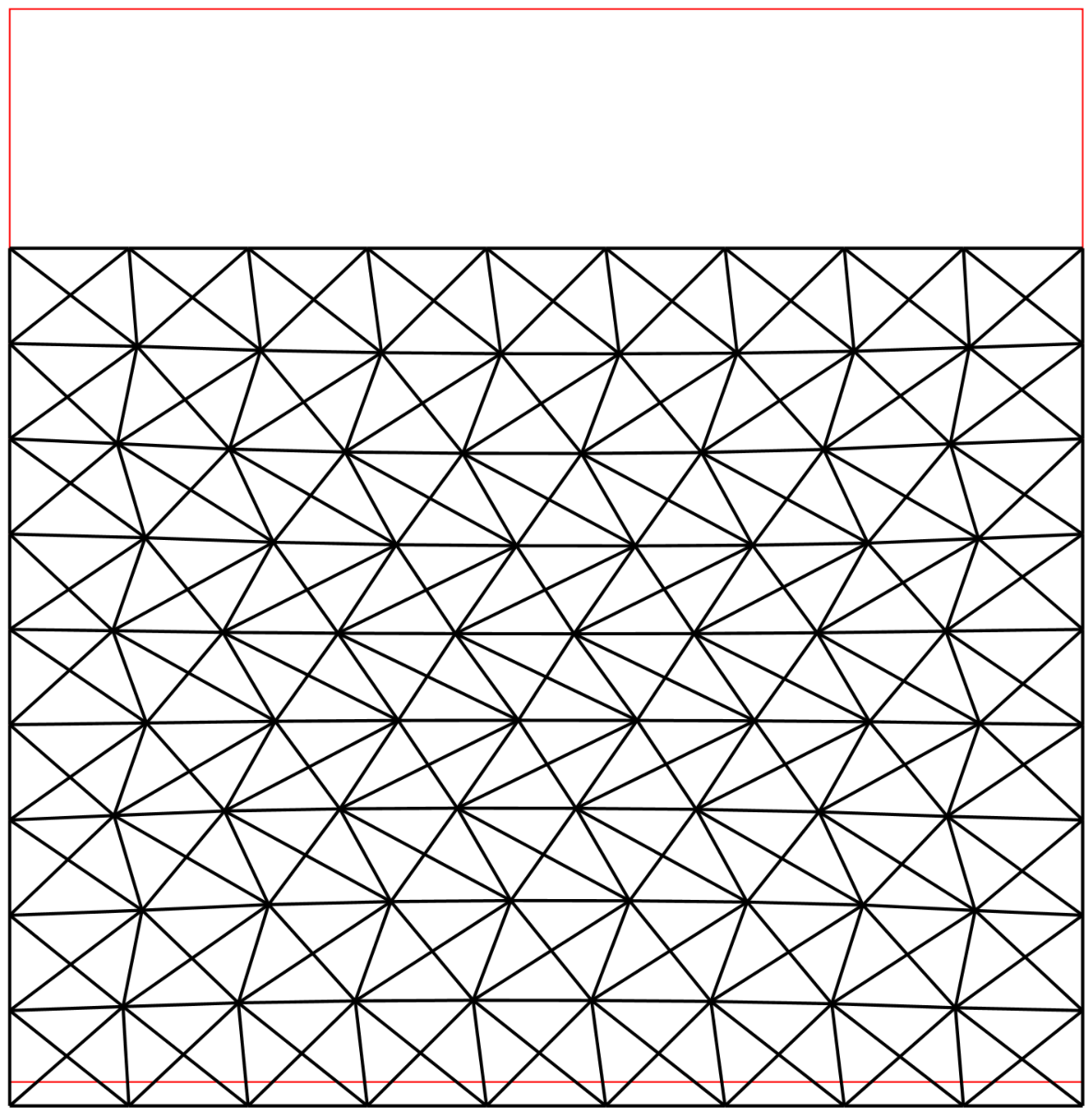}
\label{Square_high}
}
\subfigure[]{
\includegraphics[width=0.55\textwidth]{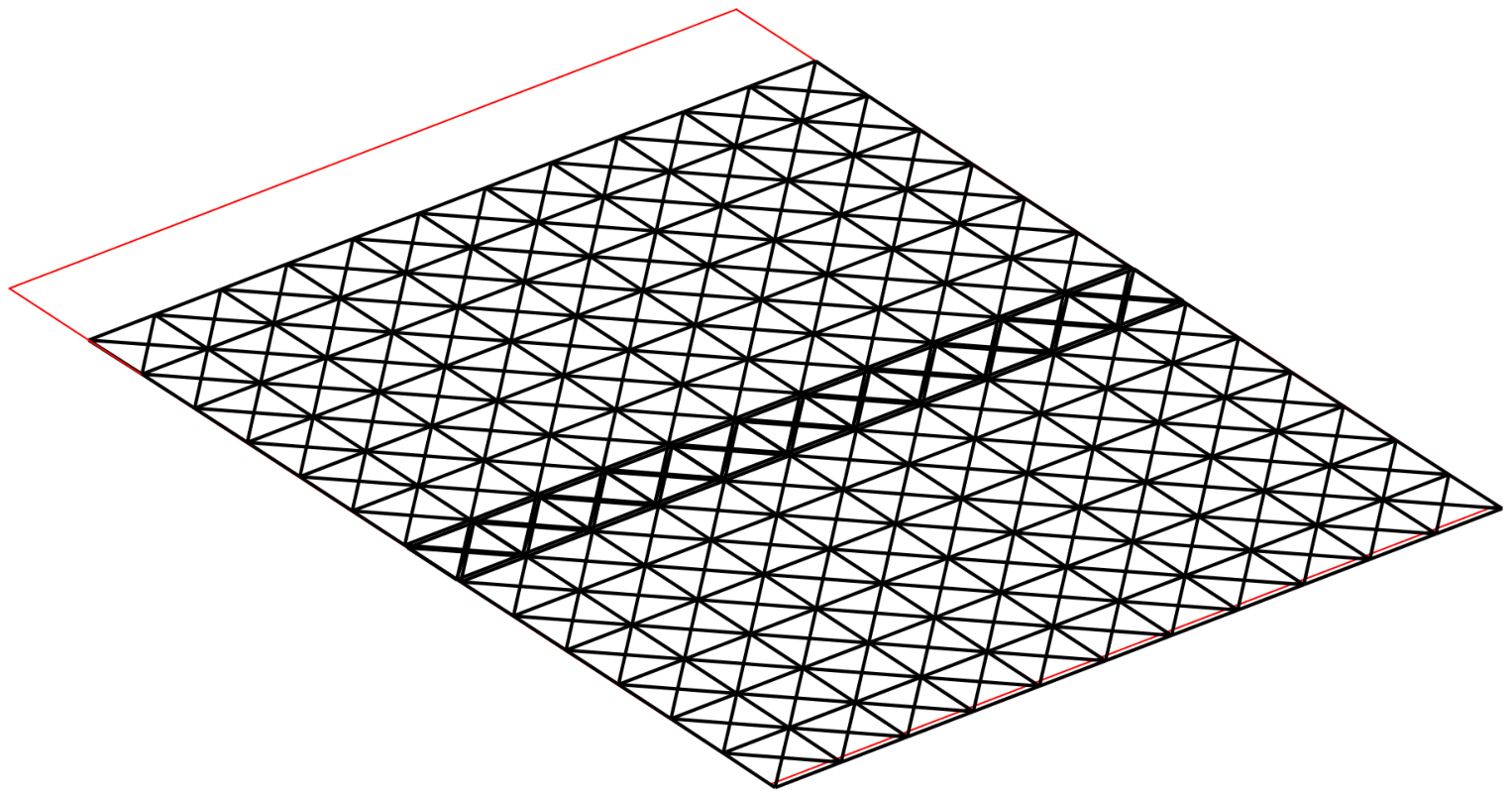}
\label{Square_low}
}
\caption{Two kinds of instabilities in square lattices subjected to compression. 
The deformation is in-plane and globally uniform at high stiffness parameter 
$\gamma=1.0$ (a) in contrast to the  out of plane and localized  deformation at 
low $\gamma=0.1$ (b).  }
\label{Square_2modes}
\end{figure}
We first investigate the quasi-static behavior of this lattice under uniaxial 
compression along $x$-direction with the strain denoted by $\delta_x$. To this 
end, we conduct numerical simulations to determine the deformed configuration by 
increasing the strain incrementally from the undeformed configuration. At each 
strain level, the equilibrium configuration is obtained by minimizing the 
potential energy $E$ of the lattice subject to the appropriate boundary 
conditions. The problem may be expressed as
\begin{equation}
\begin{aligned}
& \underset{\bX}{\text{minimize}}
& & E(\bX) = \dfrac{k_0}{2}\sum_{m = 1}^{M-1} \sum_{n=1}^{N-1} \bigl[ \left( 
\|\bx_{m,n}-\bx_{m+1,n}\| - a \right)^2 + 
																													\left( \|\bx_{m,n}-\bx_{m,n+1}\| - a \right)^2 +  \\
& & &	 \;\;\;\;\;\;\;	\left( \|\bx_{m,n}-\bx_{m+1,n+1}\| - \sqrt{2}a \right)^2 + 
																												\left( \|\bx_{m+1,n}-\bx_{m,n+1}\| - \sqrt{2}a \right)^2 \bigr] + \\
& &	&	\;\;\;\;\;\;\; \dfrac{k_g}{2} \sum_{m=1}^M \sum_{N=1}^N z_p^2 \\
& \text{subject to}
& & x_q = (1+\delta_x)X_q, \; y_q = 0  \;\;\; q = 1, \ldots, N_b, \\
& & & z_q = 0 \;\;\; q = 1, \ldots N_c. 
\end{aligned}
\end{equation}
Here $N_b$ is the total number of boundary nodes, $N_c$ is the nodes on the 
boundary along the $y$-direction, while $(X_p,Y_p,Z_p)$ and $\bx_p = 
(x_p,y_p,z_p)$ denote, respectively, the undeformed and deformed positions of 
node $p$ and $\bX=(\bx_{1,1},\bx_{1,2},\ldots,\bx_{M,N})$. Finally, $a$ is the 
distance between nearest neighbor masses in the 
lattice.

Figure~\ref{Square_2modes} illustrates 
the typical deformation field for two values of stiffness parameter $\gamma$. 
The two chosen $\gamma$ values are representative of the  behavior of lattices 
with high and low $\gamma$. Figure~\ref{Square_high} displays the in-plane 
deformation field when $\gamma = 1.0$. The out-of-plane displacement is zero and 
globally uniform patterns form after the onset of an instability. The 
displacement field is affine for small strains $\delta$ near the undeformed 
configuration. This range of $\delta$ may be determined numerically using 
a procedure similar to the stability analysis presented later in 
Sec.~\ref{firstBifSec}. With increasing strain, this affine solution loses 
stability leading to the evolution of globally uniform patterns with a 
wavelength of two unit cells along the loading direction and uniform along the 
transverse in-plane direction. We remark here that this deformation mode arises 
as  a bifurcation since the affine solution satisfies the equilibrium equations. 
Figure~\ref{Square_low} displays the deformation field for a square lattice with 
a lower stiffness parameter value $\gamma = 0.1$. Here the deformation happens 
along the out-of-plane direction too and leads to a localization which consists 
in one  layer of the lattice folding over its adjacent layer. We observe that 
the deformation field is essentially two dimensional with no displacement in the 
transverse in-plane direction ($y$). Thus we see that the patterns that form 
after the onset of a bifurcation depend on the material property, stiffness 
parameter $\gamma$ in this case, and it can range from globally uniform patterns 
to localized folding behavior.

\subsection{Deformation sequence leading to localization}

\begin{figure}
\centering
\subfigure[]{
\includegraphics[width=0.3\textwidth]{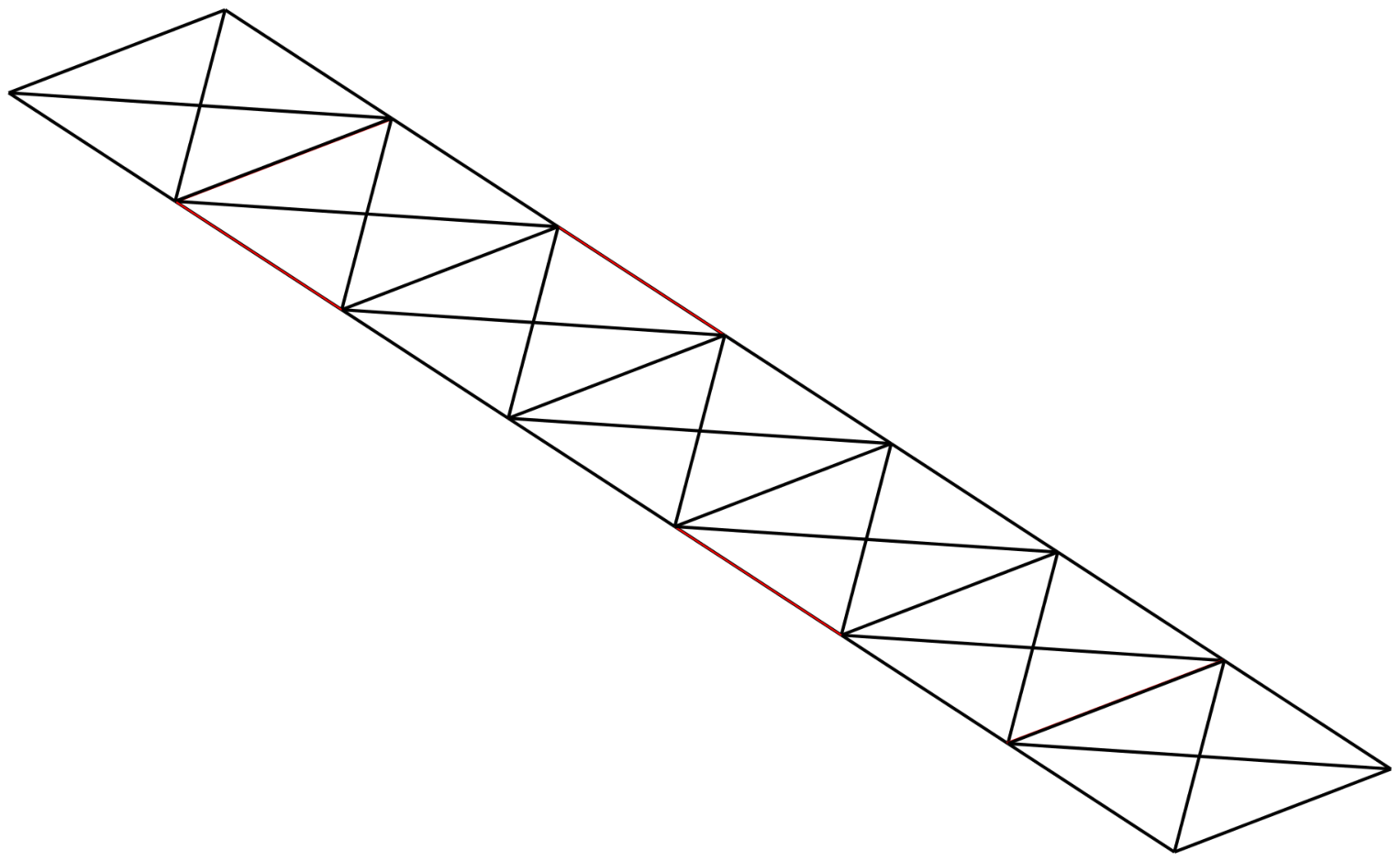}
\label{shot1}
}
\subfigure[]{
\includegraphics[width=0.3\textwidth]{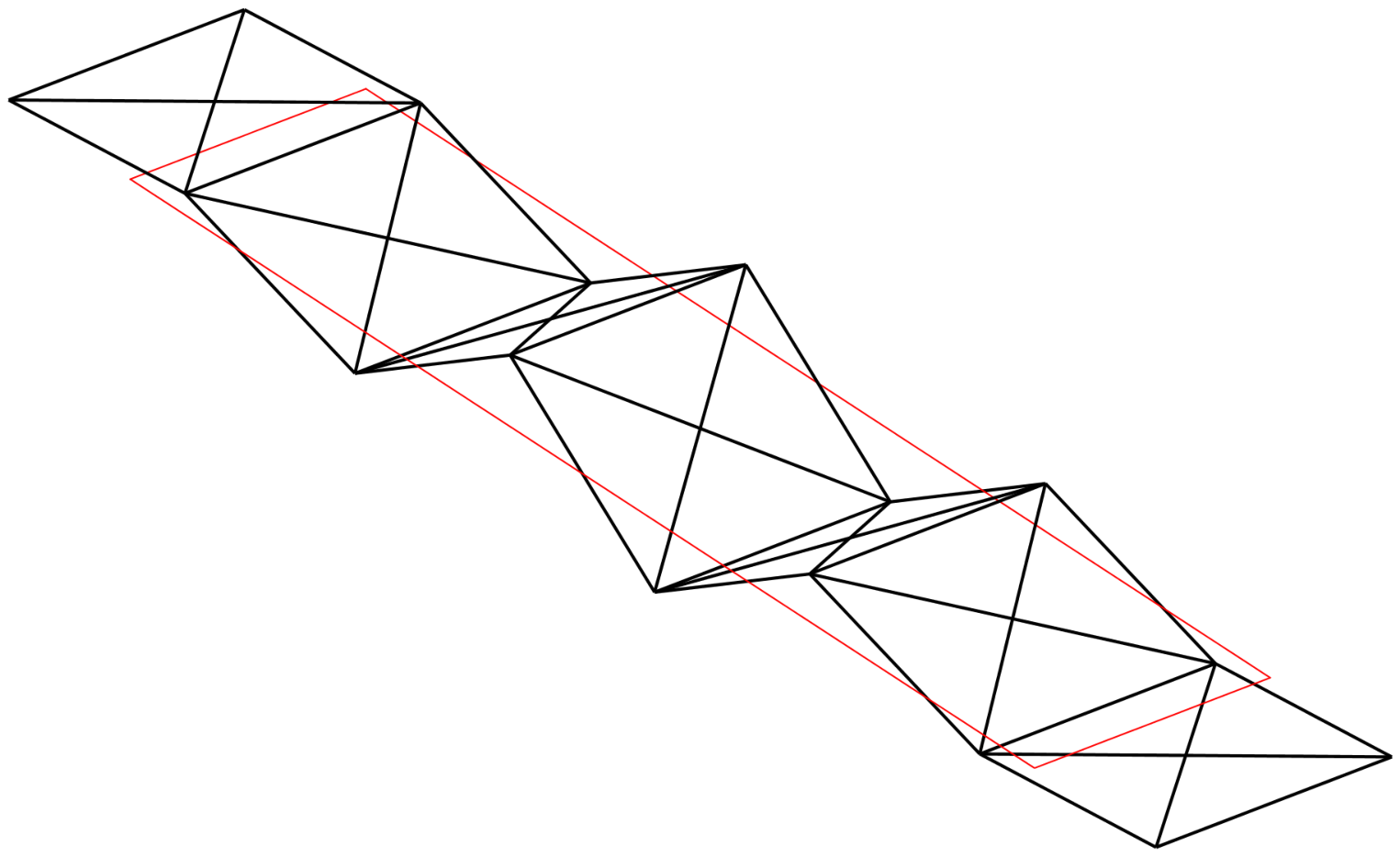}
\label{shot2}
}
\subfigure[]{
\includegraphics[width=0.3\textwidth]{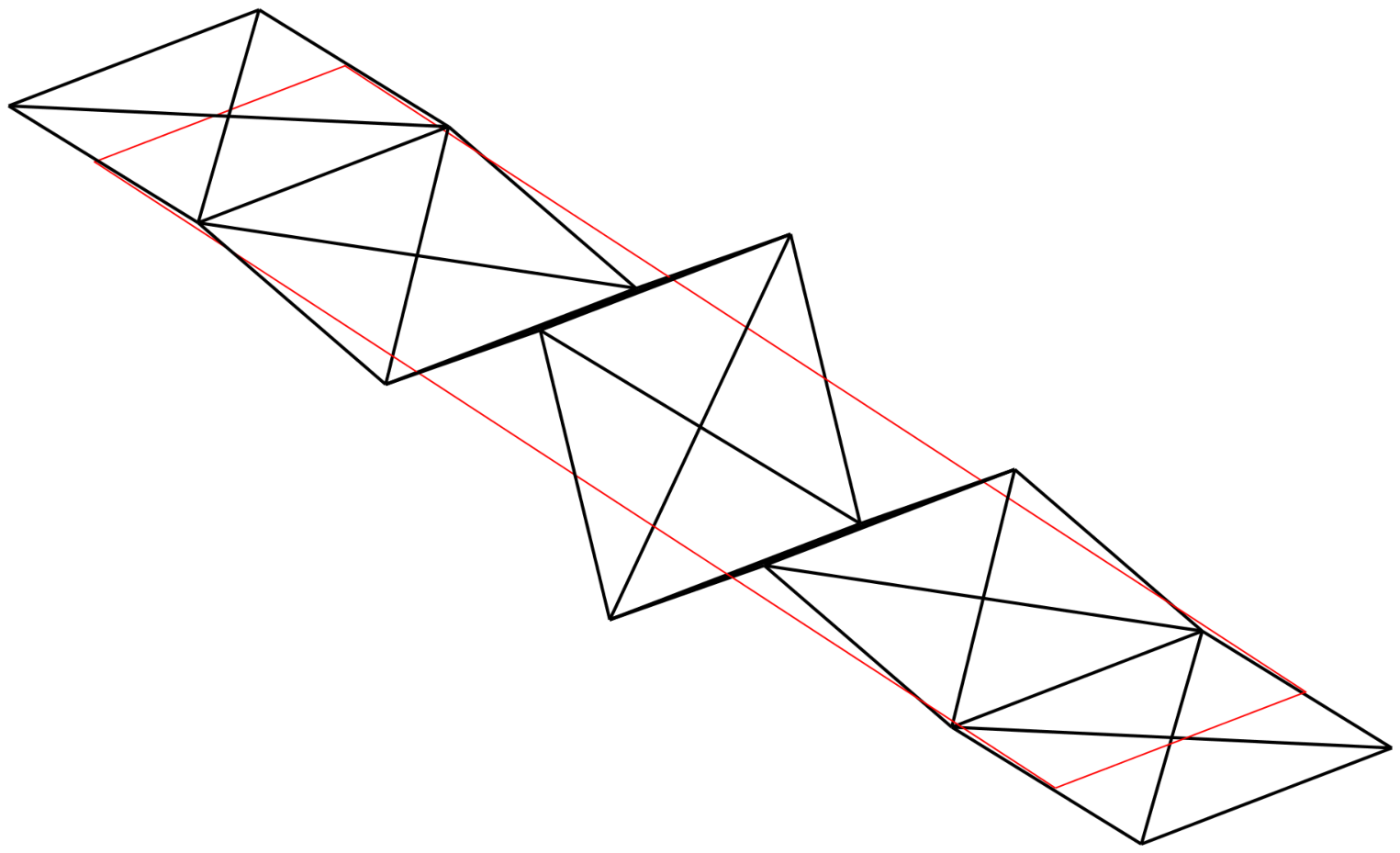}
\label{shot3}
}
\subfigure[]{
\includegraphics[width=0.3\textwidth]{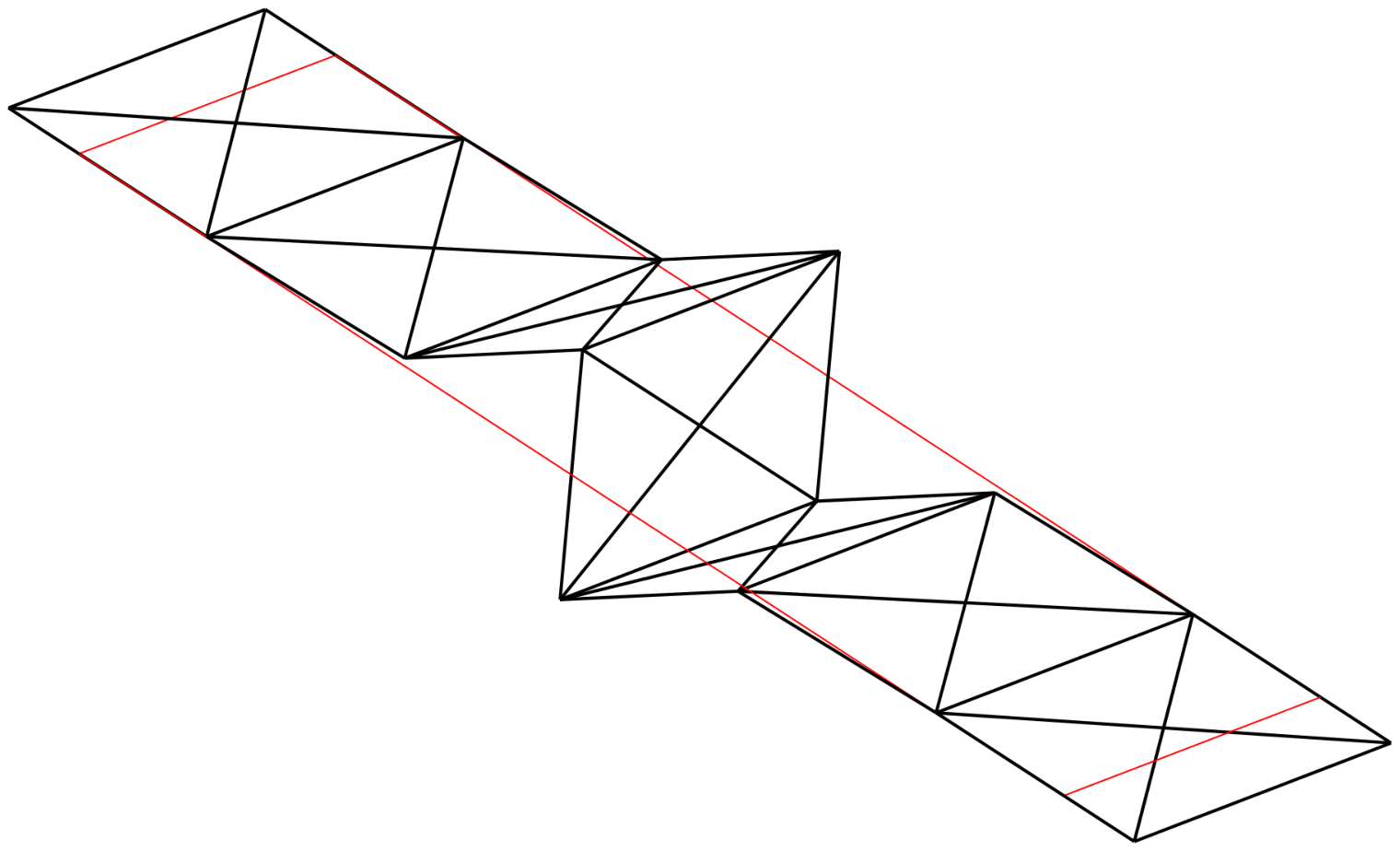}
\label{shot4}
}
\subfigure[]{
\includegraphics[width=0.3\textwidth]{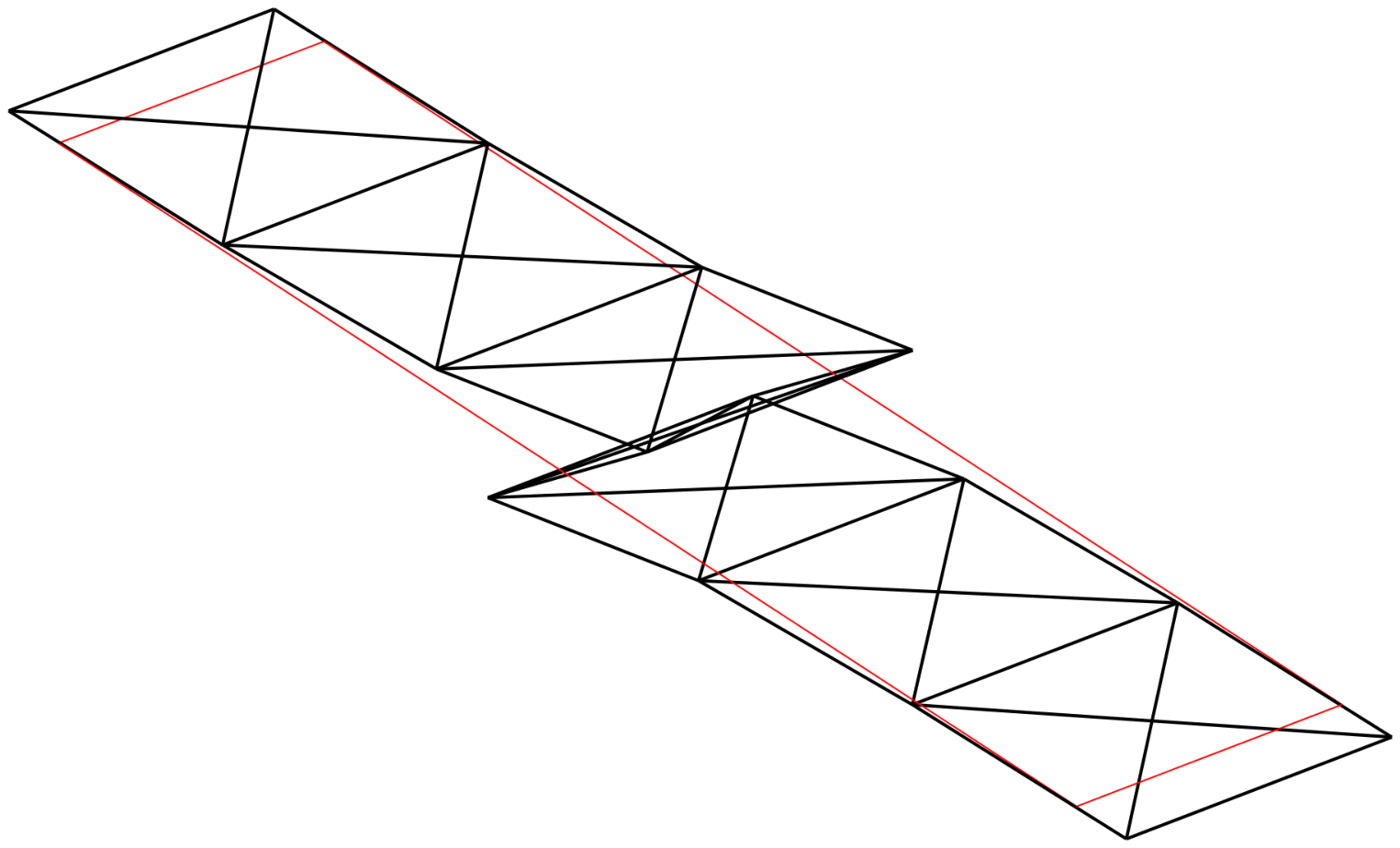}
\label{shot5}
}
\subfigure[]{
\includegraphics[width=0.3\textwidth]{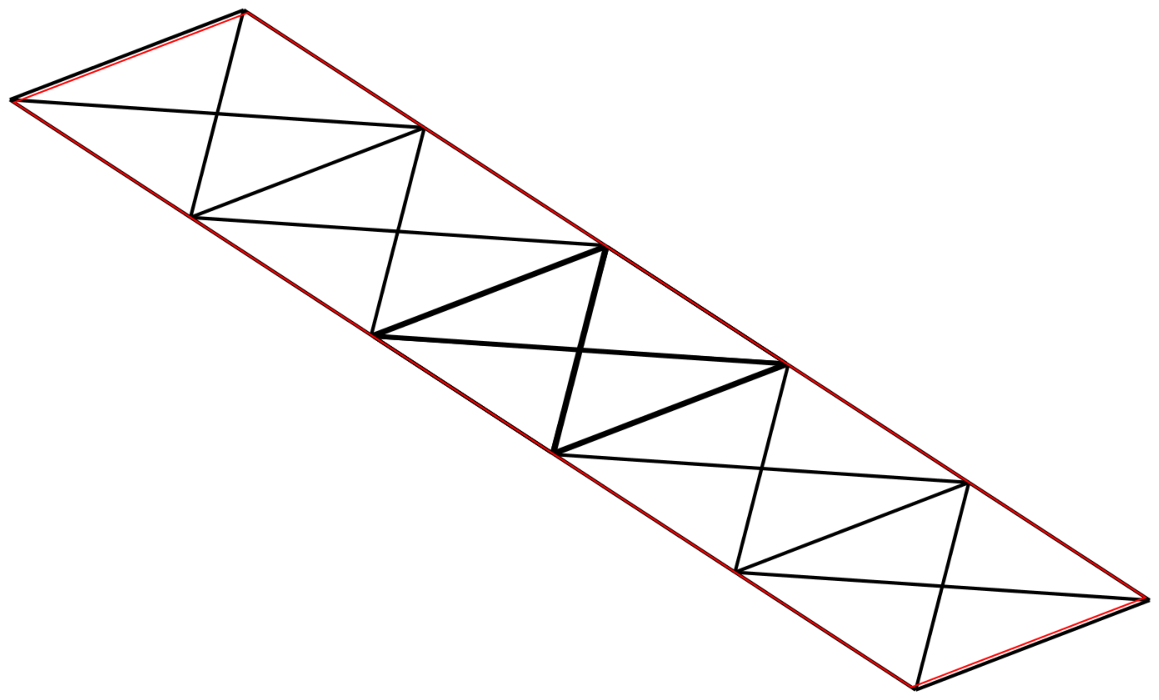}
\label{shot6}
}
\caption{Snapshots of the folding transition process from the fully expanded and 
undeformed configuration (a) to the folded configuration (f), $\gamma=0.1$. The 
process involves two distinct bifurcations, with the vertical deformation first 
increasing globally and then getting localized at the center as the lattice is 
compressed.}
\label{FoldingSnapshots}
\end{figure}
	
Let us examine the sequence of deformations that result in a localized folding 
pattern in a lattice with $\gamma=0.1$. Since there is no displacement in the 
transverse in-plane direction, we consider a single strip of the square lattice 
with $N \times M = 2 \times 8$ nodes. Figure~\ref{FoldingSnapshots} illustrates 
the sequence of deformations as this lattice is subjected to a uniaxial 
compression. For small strains close to the undeformed configuration, the 
deformation field is affine and there is no out-of-plane displacement. As the 
strain increases, this affine solution becomes unstable and the lattice deforms 
out-of-plane in a zig-zag manner (Fig.~\ref{shot2}). The out-of-plane 
displacement is maximum at the center and decreases away from it. As the strain 
increases further (Fig.~\ref{shot3}), the out-of-plane displacement increases 
for the center nodes, while it decreases for nodes away from the center. The 
out-of-plane displacement becomes localized at the two center nodes of the 
lattice and it decreases rapidly to zero away from these nodes. As the strain 
increases, the center square folds over (Fig.~\ref{shot5}-(f)), thereby 
resulting in a localized zone. Note that this localized deformation field will 
have zero energy at a strain level $\delta = -2/(M-1) \approx -0.29$ when all 
the axial springs are unstretched and the out-of-plane displacement is zero. 
The existence of multiple zero energy states (undeformed and folded 
configurations) implies  that the potential energy is not quasi-convex. It also 
points to the existence of negative stiffness regions and the possibility of 
this lattice to exhibit hysteresis. A negative effective stiffness results in a 
snap-through type behavior as the lattice transitions from the unfolded to the 
folded configuration or vice versa. We examine this rich set of behaviors 
arising due to this localized solution in the next section using a 
one-dimensional model which turns out to be quite adequate for our purposes. 

\section{Chain on elastic support}\label{chainSec}
\begin{figure}
\centering
\includegraphics[width=0.5\textwidth]{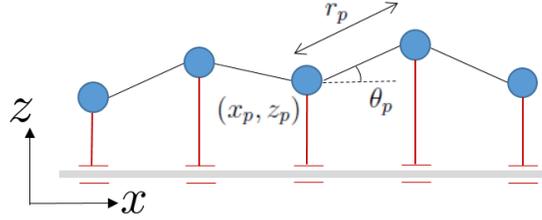}
\caption{Schematic of a $1D$ lattice model, with point masses at the nodes 
interacting with nearest neighbor horizontal springs and vertical ground 
springs. Each mass has two degrees of freedom and can move in the $xz$-plane. }
\label{shooting_schematic}
\end{figure}

\begin{figure}
\centering
\includegraphics[width=0.5\textwidth]{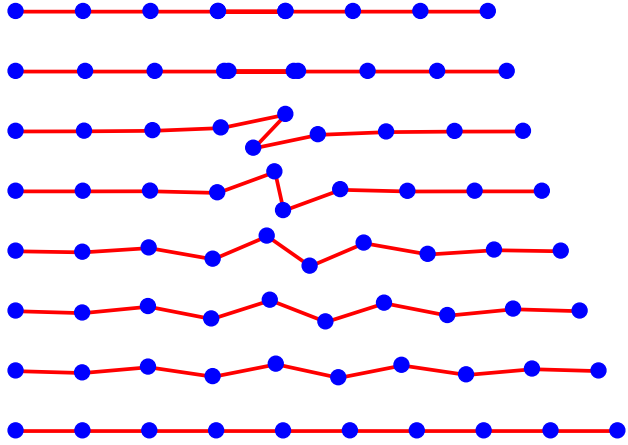}
\caption{Sequence of deformed configurations as the lattice transitions from the 
folded to the unfolded configuration.} 
\label{beam_schematic}
\end{figure}

To understand the mechanism of the snap-through behavior during localization, 
let us further simplify the square lattice and consider an equivalent $1D$ 
(one-dimensional) model obtained by taking one section of the square lattice 
deprived of the diagonal springs. Figure~\ref{shooting_schematic} displays a 
schematic of the considered lattice model. There are $N$  nodes along a strip 
indexed from $1$ to $N$. The potential energy $E$ of the lattice may be 
written as 
\begin{equation}\label{pot}
E(\bX) = 		\sum_{i=1}^{N}\dfrac{k_g}{2}  z_i^2 + 
		\sum_{i = 1}^{N-1} \dfrac{k_0}{2}  \left(\|\bx_{i} - 
\bx_{i+1}\|- a\right)^2 . 
\end{equation}
where $\bx_i=(x_i,z_i)$ and $\bX=(\bx_1,\bx_2,\ldots,\bx_N)$. Here we 
can assume that the undeformed axial springs have unit length ($a=1$). 
Minimizing this energy subject to the boundary conditions yields the equilibrium 
configuration. We illustrate the typical displacement response history as the 
folding transition happens in this model, see Fig.~\ref{beam_schematic}. 

We first conduct numerical simulations on a lattice of $N=10$ masses by 
subjecting it to both a compressive strain from the unfolded configuration and a 
tensile strain from the folded configuration. For both cases, we denote the 
strain by $\delta$, measuring it from the reference unfolded configuration. The 
numerical simulations are again conducted using a Newton-Raphson solver. This 
numerical solver based on incrementally applying strain faces convergence issues 
in the presence of instabilities and bifurcations. To overcome these 
limitations, we develop alternative techniques in Sec.~\ref{transitionSec}.

\subsection{Folding behavior}

This $1D$ model also undergoes a folding based localization and we will analyze 
the deformation process leading to this localization in detail. In all our 
subsequent calculations, we consider a chain of 10 point masses (nodes) 
connected by axial springs of stiffness $k_0$. Similar to the square lattice, 
each mass is also connected to the ground by a spring of stiffness $k_g$ and the 
quasi-static behavior is then governed by the non-dimensional stiffness 
parameter $\gamma = k_g/k_0$. The chain starts from $x=0$ and the undeformed 
lattice spans from $x_1 = 0$ to $x_{10} = 9$. This is referred to as the 
unfolded configuration. The first mass is kept fixed while the last mass is 
subjected to a compressive displacement. The two end masses are also restricted 
from moving vertically ($z_1=z_{10} =0$). As the prescribed compressive 
displacement increases, the lattice folds at the center and now spans from $x_1 
= 0$ to $x_{10} = 7$. This is referred to as the folded configuration. Thus we 
have that $\delta=0$ for the unfolded configuration while $\delta=-2/9$ for the 
folded one. Figure~\ref{beam_schematic} displays snapshots of the deformed 
lattice as it deforms from one configuration to the other. We will analyze the 
quasi-static response of the lattice as it deforms from the folded to the 
unfolded configuration and vice versa. 

\begin{figure}
\centering
\subfigure[]{
\includegraphics[width=0.45\textwidth]{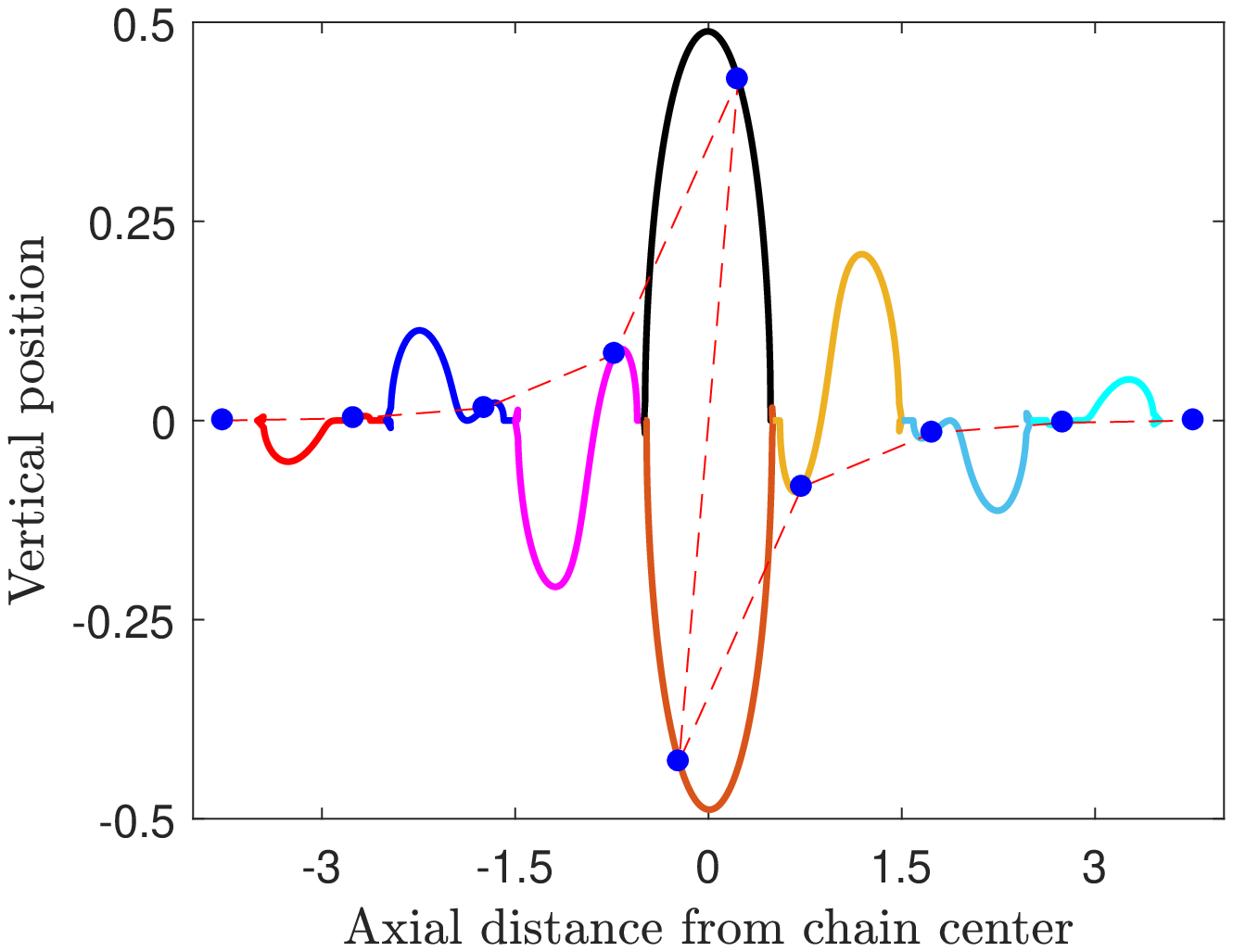}
\label{zdisp_low1}
}
\subfigure[]{
\includegraphics[width=0.45\textwidth]{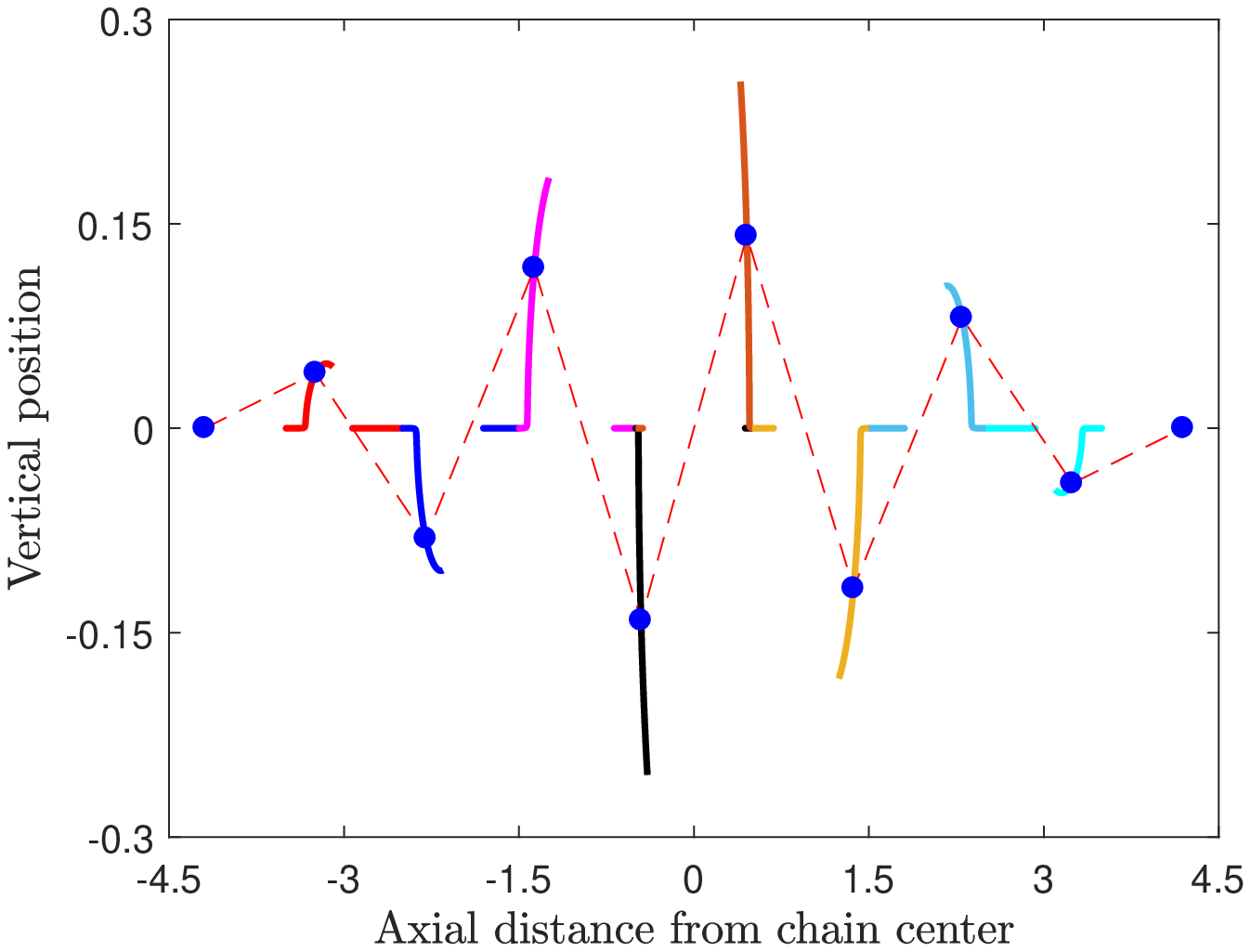}
\label{zdisp_high1}
}
\subfigure[]{
\includegraphics[width=0.45\textwidth]{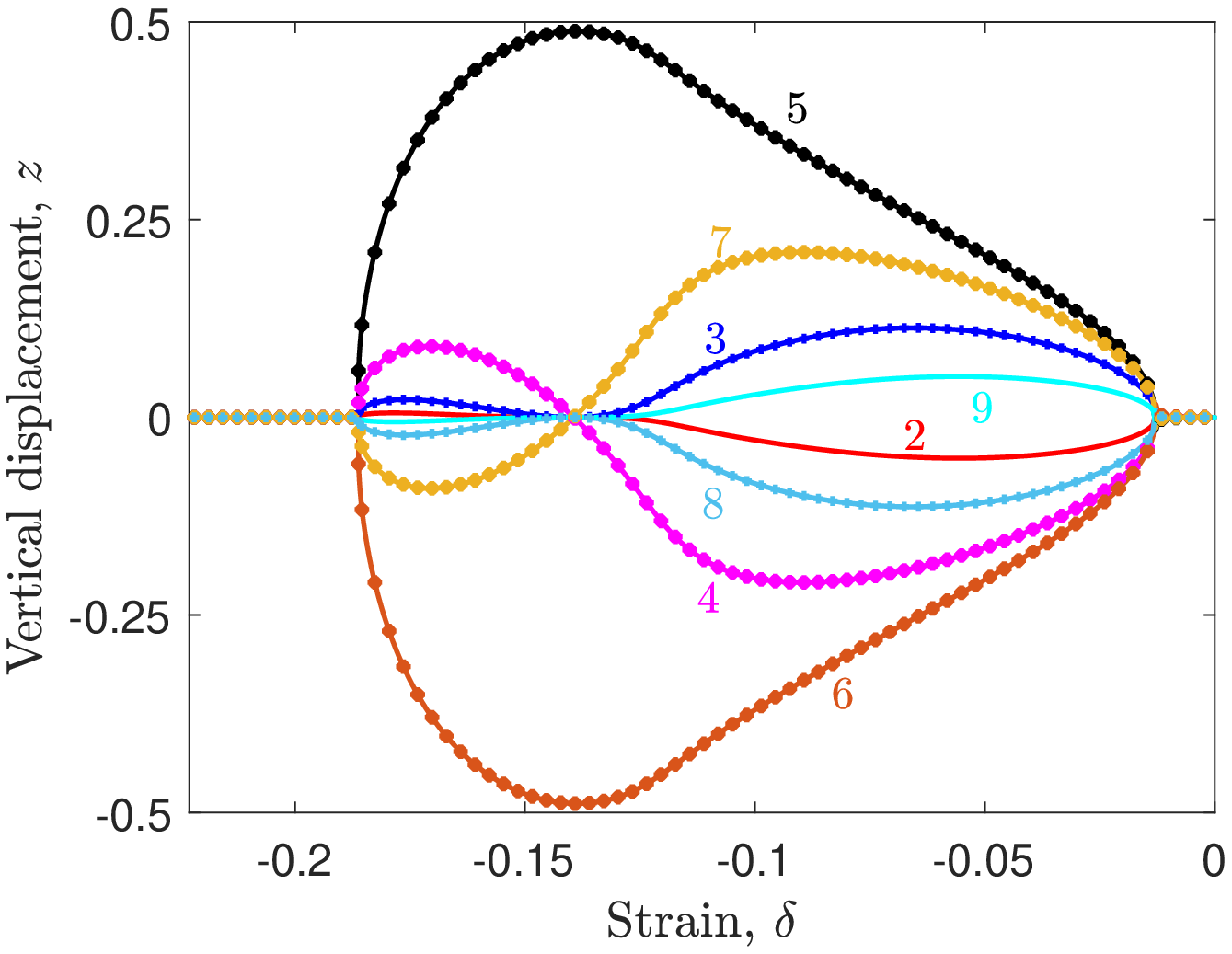}
\label{zdisp_low}
}
\subfigure[]{
\includegraphics[width=0.45\textwidth]{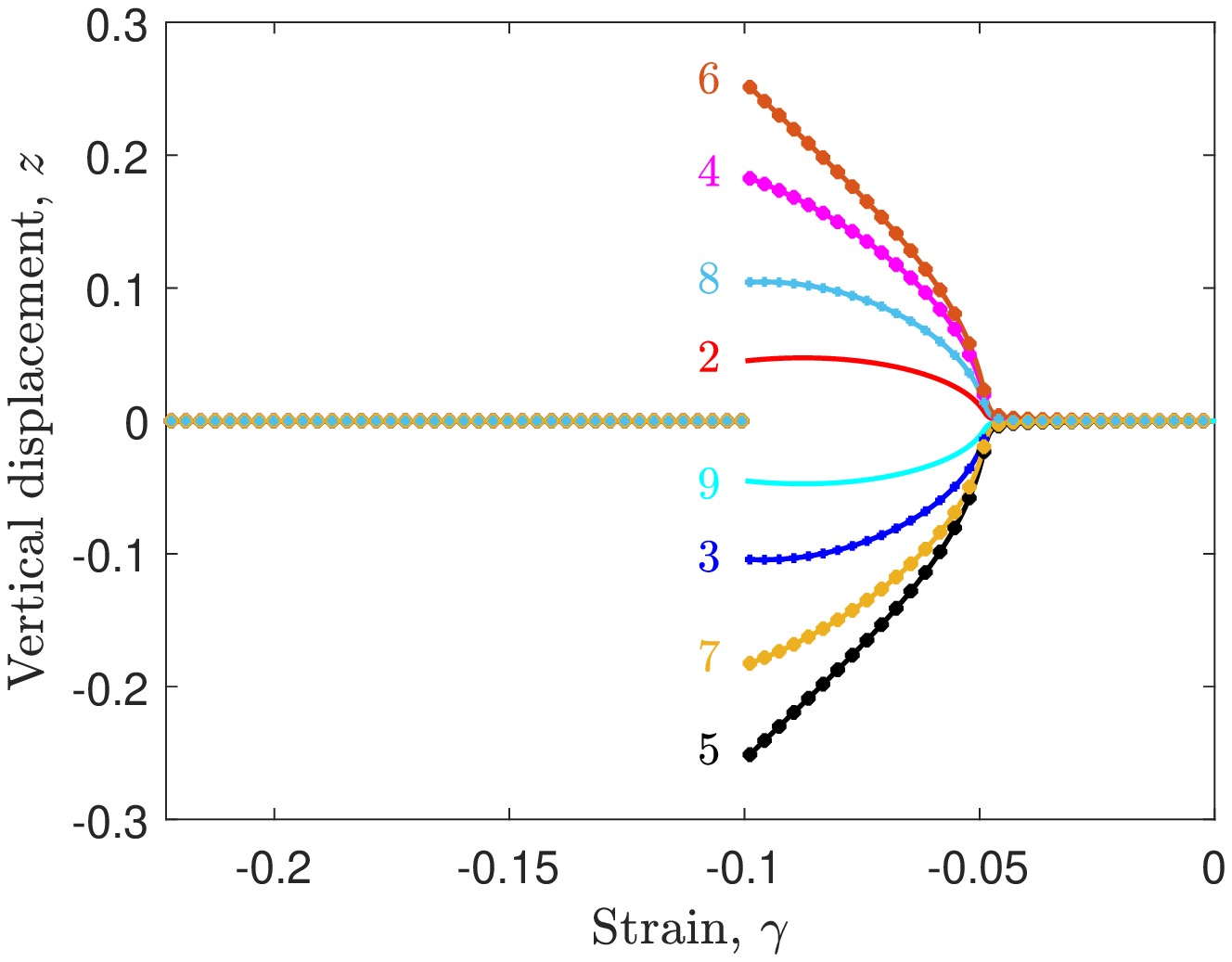}
\label{zdisp_high}
}
\caption{Vertical displacement of the nodes as the lattice is pulled from the 
folded configuration. Two bifurcation points are observed at low stiffness 
parameter $\gamma = 0.05$ (a, c), while a single bifurcation and a jump 
in displacement is obtained at high $\gamma=0.2$(b, d).  }
\label{zdispNodes}
\end{figure}

Figure~\ref{zdispNodes} displays the vertical displacement of the nodes as this 
lattice is subjected to a tensile strain from the folded configuration. Note 
that the lattice in the initial configuration spans from $x_1 = 0$ to $x_{10} = 
7$ in the axial direction. As the lattice is pulled from both ends, it reaches a 
final unfolded and undeformed configuration ranging from $0$ to $9$. The 
horizontal and vertical axes show, respectively, the compressive strain  
$\delta$ measured from the reference unfolded configuration and the vertical 
displacement of all the nodes in between. Figure~\ref{zdisp_low} displays the 
displacements for  stiffness parameter $\gamma = 0.1$. At small displacements 
from the initial configuration, the axial displacement field is affine in the 
lattice and the vertical displacement is zero. At around $\delta = 
-0.187$, there is an instability and the lattice deforms out of plane, thereby 
initiating the unfolding process. The vertical displacement decreases away from 
the center nodes and it increases with increasing tensile strain near the 
bifurcation point. The nodes to the right of the center spring have 
a positive vertical displacement while the corresponding nodes to the left have 
an equal and opposite vertical displacement.

As the tensile strain increases, the vertical displacement of the nodes away 
from the center decrease to zero at around $\delta=-0.139$. Beyond this point, 
the shape of the displacement field changes as the vertical displacement of 
adjacent nodes alternate in sign. They first increase and then decrease in 
magnitude, reaching a zero value at around $\delta = -0.011$. Beyond this point, 
the displacement field is affine in the axial direction and has no vertical 
component. This deformation mode corresponds to uniform compression of the 
lattice from the unfolded and undeformed position, which is attained at $x_{10} 
= 9$. We thus see two distinct mode-shapes arising from the onset of two 
bifurcation points as the lattice deforms from the folded to the unfolded 
configuration. We address the mechanism of transformation from 
one mode-shape to another using a reduced $4$-mass model later in 
Sec.~\ref{4masschain}.

Let us now examine the displacement when a lattice with a higher stiffness 
parameter value $\gamma=0.2$ is subjected to a similar tensile strain. 
Figure~\ref{zdisp_high} displays the vertical displacement of the interior nodes 
($z_2$ to $z_9$)  as it deforms from the folded to the unfolded configuration. 
In contrast to the low stiffness parameter case, here the vertical displacement 
remains zero for larger tensile strains until around $\delta=-0.1$, when 
the vertical displacement jumps sharply to a non-zero value. The vertical 
displacement of the adjacent nodes has opposite sign and resembles the second 
mode-shape discussed above. As the tensile strain increases, the vertical 
displacement decreases and reaches zero at around $\delta = -0.044$. Beyond this 
point, the vertical displacement remains zero and the axial displacement is 
affine with equal compressive strain in each of the axial springs. 

\subsection{Linear stability analysis}\label{firstBifSec}

To investigate the local stability of the equilibria computed above, we need to 
look at the Hessian $\bH$ of the energy $E(\bX)$, in \eqref{pot}. For a chain 
having $N$ masses it can be expressed as
\begin{equation}
\bH = \bI_N  \otimes \begin{pmatrix} 0 & 0 \\ 0 & \gamma  \end{pmatrix} + 
\sum_{p=1}^{N-1}\bK_{h,p},
\end{equation}
where $\bI_N$ is the $N\times N$ identity matrix and $\otimes$ denotes 
the tensor product. Here $\bK_{h,p}$ is the contribution due to the $p$-th 
axial spring having a deformed length $r_p$, with an angle $\theta_p$ with 
respect to the horizontal axis, see Fig.~\ref{shooting_schematic}. It can be 
written as
\begin{equation}
\bK_h = \bD_p \otimes (\bR_p \bK_{loc,p} \bR_p^T).
\end{equation}
where the $\bD_p$ is a $N\times N$ matrix what all 0 entries but for the 
$(p,p+1)$ principal minor,
\[
 \bD_p=\bordermatrix{ & & & p & p+1 & & \cr
                      & 0 & \cdots & \cdots & \cdots & \cdots & 0 \cr
                      & \vdots & \ddots & &  & \vdots & \vdots\cr
                    p & 0 & \cdots & 1 & -1 & \cdots & 0\cr
                  p+1 & 0 & \cdots &-1 & 1 & \cdots & 0\cr
                      & \vdots & \vdots & & \ddots & \vdots \cr
                      & 0 & \cdots & \cdots & \cdots & \cdots & 0}\ ,
\]
while $\bR_p$ is the rotation of angle $\theta_p$ and $\bK_{loc,p}$ is related 
to the stiffness of the axial spring in the local coordinate system
aligned with the spring axis:
\begin{equation}
\bR_p = \begin{pmatrix} \cos \theta_p & -\sin \theta_p \\ \sin \theta_p & \cos 
\theta_p \end{pmatrix} , \qquad
\bK_{loc,p} = \begin{pmatrix} 1 &  0 \\ 0 & 1 - \dfrac{1}{r_p} \end{pmatrix} 
\;\;\; .
\end{equation}
It is easy to observe that if all the springs are horizontal, that if 
$\theta_p=0$ or $\pi$ for every $p$, then the $x$ and the $z$ part of the 
Hessian 
$\bH$ decouple. More precisely, let $\bP$ be the permutation matrix that send 
$\bX$ to $\widetilde \bX=(x_1,x_2,\ldots, x_N,z_1,z_2,\ldots,z_N)=(X,Z)$ then 
we get
\[
 \widetilde\bH=\bP\bH\bP^T=\begin{pmatrix}\bH_x & 0\\0 & \bH_z\end{pmatrix}.
\]
Moreover one can verify that the $x$ part of the Hessian $\bH_x$ is 
always positive definite. Thus to analyze the stability of an equilibrium with 
no displacement in the $z$ direction we just need to look at the eigenvalues of 
$\bH_z$.

Let us consider a lattice compressed from its initial unfolded configuration to 
a strain $\delta$. The affine deformation induced by such a strain is 
given by $x_p=(1+\delta)p$ and $z_p=0$ for all $p$. This is clearly 
an equilibrium configuration. Since the first and last mass are fixed, the  
stability of this configuration is controlled by the eigenvalues of the 
$(N-2)\times(N-2)$ matrix
\begin{equation}\label{Hess}
 \bH_z^u=\bordermatrix{&2 & 3 & & & N-2\cr
2 & \alpha & \zeta & 0 & \cdots & 0 \cr
3 & \zeta & \alpha & \zeta & \cdots & 0 \cr 
\vdots & \ddots & \ddots & \ddots & \vdots  \cr 
N-3 & 0 & \cdots & \zeta &  \alpha & \zeta \cr
N-2 & 0 & \cdots & 0 &  \zeta & \alpha },
\end{equation}
where 
\begin{equation}
\alpha = \gamma + \dfrac{2\delta}{1+\delta} , \;\;
\zeta =-\dfrac{\delta}{1 + \delta}. 
\end{equation}
By looking for eigenvectors $X_p$ of the form $X_{p,q}=\sin(\omega_p (q-1))$, 
$q=2,\ldots,N-1$, one immediately gets $\omega_p=\pi p/(N-1)$ with associated 
eigenvalue
\[
 \lambda_p=\gamma+\frac{2\delta}{1+\delta}\left(1-\cos(\omega_p)\right)
\]
and $p=1,\ldots,N-2$. Thus at the critical strain 
\[
 \delta^*_u = \frac{-\gamma}{\gamma+2\left(1-\cos(\omega_{N-2})\right)}
\]
$\lambda_{N-2}$ becomes negative and the affine configuration looses stability.
The associated mode shape is given by
\[
X_{N-2,q}=\sin\left(\frac{(N-2)(q-1)\pi}{N-1}\right)
\]
that shows a maximum value at the center which explains the high deformation, 
which leads to subsequent folding at the center. The strain value and mode 
shape predicted by the above stability analysis is in excellent agreement with 
the numerical solution.

\begin{figure}
\centering
\includegraphics[width=0.6\textwidth]{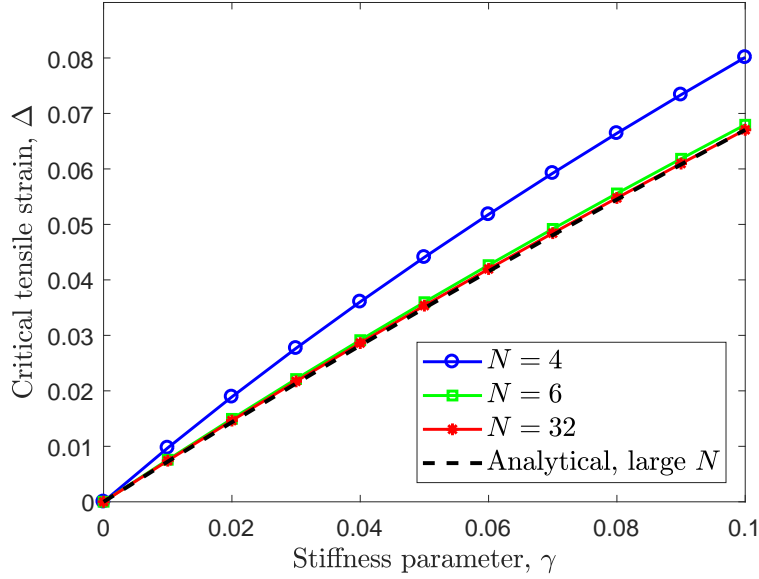}
\caption{Tensile strain in the axial spring with stiffness parameter $\gamma$ at 
the onset of unfolding for three different chain lengths. A chain of six masses 
can essentially capture the onset of bifurcation from the folded configuration 
of a much large chain with reasonable accuracy. }
\label{Unfolding_ak}
\end{figure}

Let us now consider what happens when a folded lattice is stretched. The folded 
reference configuration is given by $z_p=0$ for all $p$ while $x_p=p-1$ if 
$p\leq N/2$ and $x_p=p-3$ if $p> N/2$. Thus the strain of this configuration is 
$\delta=-2/(N-1)$. If we stretch this configuration and look for an 
equilibrium still having $z_p=0$ for all $p$ we need all springs but the 
central folded one to have the same strain $\Delta$ while the central one has 
strain $-\Delta$. We thus get $x_p=(p-1)(1+\Delta)$ if $p\leq N/2$ while 
$x_p=(p-2)(1+\Delta)-(1-\Delta)$ if $p> N/2$. The total strain $\delta$ of this 
configuration is thus $\delta=\Delta-2/(N-1)$. Again we want the critical value 
of $\Delta$, and thus $\delta$, for which this configuration looses stability.

The Hessian $\bH^f_z$ in this configuration is similar to \eqref{Hess}, that 
is 
\begin{equation}\label{stabfold}
  \bH^f_z=\begin{pmatrix}
\alpha_2 & \zeta_2 & 0 & \cdots & 0 \\ 
\zeta_2 & \alpha_3 & \zeta_3 & \cdots & 0 \\ 
\vdots & \ddots & \ddots & \ddots & \vdots  \\ 
0 & \cdots & \zeta_{N-3} &  \alpha_{N-2} & \zeta_{N-2} \\
0 & \cdots & 0 &  \zeta_{N-2} & \alpha_{N-1} 
       \end{pmatrix},
\end{equation}
where
\[
 \alpha_p=\gamma+\frac{2\Delta}{1+\Delta}\qquad \zeta_p=-\frac{\Delta}{1+\Delta}
\]
for all $p$ but for
\[
\alpha_{\frac{N}{2}}=\alpha_{\frac{N}{2}+1}=\gamma+\frac{\Delta}{1+\Delta}-\frac
{\Delta}{ 1-\Delta} \qquad \zeta_{\frac{N}{2}}=\frac{\Delta}{1-\Delta}\ .
\]
Let $\bQ$ be the reflection matrix such that $(\bQ X)_p=X_{N-p}$. Clearly 
$\bH^f_z$ commutes with $\bQ$ so that we know all eigenvectors $X$ of 
$\bH^f_z$ must satisfy $\bQ X=\pm X$.  This means that they are either 
symmetric or anti-symmetric with respect to the reflection $p\to N-p$. Taking 
this into consideration, we find that the search for the eigenvalues of 
$\bH^f_z$ can be 
reduced to that of the eigenvalues of the two $(N/2-1)\times (N/2-1)$ matrices 
given by

\begin{equation}\label{Heqn}
\bH_\pm = \begin{pmatrix} \alpha & \zeta & & &  \\ 
\zeta & \alpha & \zeta & 0 &  \\ & & \ddots & &  \\ & 0 & &  \alpha & \zeta \\
& & &  \zeta & \beta_\pm   \end{pmatrix}, 
\end{equation}
where 
\begin{equation}
\alpha = \gamma + \dfrac{2\Delta}{1+\Delta} , \qquad \zeta =-\dfrac{\Delta}{1 + 
\Delta} 
\end{equation}
and 
\[
 \beta_+=\gamma + \dfrac{\Delta}{1+\Delta}\qquad 
\beta_-=\gamma + \dfrac{\Delta}{1+\Delta}-\dfrac{2\Delta}{1-\Delta}\ .
\]
Here $\beta_+$ refers to the symmetric eigenvalues while $\beta_-$ to the 
anti-symmetric ones.

As before it is natural to look for 
eigenvectors $X_p$ of the form $X_{p,q}=\sin(\omega_p (q-1))$, 
$q=2,\ldots,N/2-1$. This leads to the condition
\[
 (\beta_{\pm}-\alpha)\sin(\omega M)=\zeta \sin(\omega (M+1))
\]
where $M=N/2-1$.

In the symmetric case this reduces to
\[
\sin(\omega M)=\sin(\omega (M+1))
\]
so that we get the solutions $\omega_p=\pi p/(2M+1)$ with 
eigenvalue $\lambda_p=\alpha+2\gamma\cos(\omega_p)$. It is easy to see that 
these eigenvalues are always positive for $\Delta>0$. In the anti-symmetric 
case we get
\begin{equation}\label{eig}
 \sin(\omega(M+1))=\dfrac{3+\Delta}{1-\Delta}\sin(\omega M).
\end{equation}
Observe that since 
\[
 \dfrac{3+\Delta}{1-\Delta}>\dfrac{M+1}{M}
\]
then \eqref{eig} admits also exactly one complex solution $\omega_r=i\nu^r$. 
The associated eigenvector is thus $X^r_q=\sinh \nu^r (q-1)$ with eigenvalue 
$\lambda^r=\alpha+2\gamma\cosh(\nu^r)$. A complete basis of eigenvectors is 
then obtained from the real solution of \eqref{eig}. It is not hard to see that 
if $\omega$ is real than the associated eigenvalue is positive for $\Delta>0$.

It is not easy to give an exact expression for $\nu^r$ for finite $r$. If we 
take the limit $M\to \infty$ and assume that $\nu^r$ has a limit, we obtain, 
after some algebra,
\[
\cosh(\nu^r)=\frac12\left(\frac{3+\Delta}{1-\Delta}+
\frac{1-\Delta}{3+\Delta}\right)=\frac{5+2\Delta+\Delta^2}{3-2\Delta 
-\Delta^2}
\]
so that 
\[
 \lambda^r=\frac{ (\gamma +4)\Delta^2+2  (\gamma +2)\Delta-3 
\gamma}{\Delta^2+2\Delta-3}
\]
We thus obtain that the critical strain at which the folded configuration 
loses stability is given by
\begin{equation}
 \Delta^*=\frac{-(\gamma +2)\pm2\sqrt{(\gamma+2)^2-3}}{\gamma+4}.
\end{equation}

To evaluate how close the above asymptotic value of $\Delta^*$ is to the real 
value for finite $N$ we computed $\Delta^*$ numerically. Since the eigenvalues 
of $\bH_-$ in \eqref{Heqn} are distinct, only one can become zero at the onset 
of the instability and thus the stability of this configuration can be evaluated 
by simply examining the sign of the determinant of the matrix $\bH$. 

Figure~\ref{Unfolding_ak} displays the critical tensile strain $\Delta^*$ at 
the onset of the unfolding bifurcation transition for a range of stiffness 
parameter values. Three distinct chain lengths are considered, with $N = 4, 6$ 
and $32$ point masses. In each case, the critical spring length increases almost 
linearly with the stiffness parameter $\gamma$. Furthermore, this critical value 
in a finite chain converges rapidly to the corresponding value in an infinite 
chain. Indeed, the critical strain in a chain with $6$ masses is quite close to 
that in the $32$ mass chain. 

\subsection{Unfolded to folded transitions}\label{transitionSec}
Having clarified the folding transition and identified the onset of 
bifurcation points using a stability analysis, let us now turn attention to 
analyzing the transition path between the two zero energy configurations. We use 
two numerical techniques for this purpose. We first develop and use a shooting 
method based procedure to investigate this folding transition in detail in 
Sec.~\ref{shootingSec}. We use this method to illustrate the stability regions 
as $\gamma $ increases and the associated hysteretic behavior. For a moderate 
number of masses, 10 in our present study, we 
found this approach hereafter outlined useful in understanding the bifurcation 
structure. We are aware that this approach may have limitations 
for large number of masses. We then use a classical arc length solver 
based continuation scheme in Sec.~\ref{4masschain} on a further simplified four 
mass chain which confirms the folding transition.

\subsubsection{Shooting method for ten mass chain}\label{shootingSec}

Here we cast the governing equations for the equilibria into the 
evolution of a dynamical system. Note that the governing equation of an interior 
node $p$ depends only on its nodal location and the location of its nearest 
neighbor nodes ($p-1$ and $p+1$). The governing equations may be used to solve 
for the nodal locations of $p+1$ node if the other two nodal locations are 
provided. Accordingly, we may write a relation of the form
\begin{equation}\label{dyna}
f(\bx_{p+1})=f(\bx_p,\bx_{p-1})
\end{equation}
where $\bx_p=(x_p,z_p)$. To determine explicit expressions for $\bx_{p+1}$ 
above, as before we let
$x_{p+1}-x_{p} = r_{p}\cos\theta_{p}$ and $z_{p+1}-z_{p} = 
r_{p}\sin\theta_{p}$. Figure~\ref{shooting_schematic} displays a schematic 
of a part of the chain with the relevant variables. 

With this setup, the equilibrium equations for the mass $p$, obtained by 
resolving the forces due to the springs in the horizontal and vertical 
directions, may be written as
\begin{subequations}
\begin{align}
 (r_{p} -1) \cos\theta_{p} &= (r_{p-1} - 1)\cos \theta_{p-1}, \\
 (r_{p} -1) \sin\theta_{p}  &= (r_{p-1} - 1)\sin \theta_{p-1} + \gamma z_p
\end{align}
\end{subequations}
These equations can be unambiguously solved to obtain $r_p$ and $\theta_p$ as a 
function  of $\bx_{p-1}$ and $\bx_p$. An explicit formula for $\bx_{p+1}$ is 
then easily obtained using
\begin{equation}
x_{p+1} = x_p + r_{p}\cos\theta_{p}, \;\; z_{p+1} = z_p + 
r_{p}\sin\theta_{p}.
\end{equation}

As discussed in Sec.~\ref{firstBifSec}, the onset of instability 
for both the folded and unfolded configurations happens along an anti-symmetric 
mode shape with respect to the reflection $\bQ$, see discussion after 
\eqref{stabfold}. We thus expect that the deformation process from the 
unfolded to the folded configuration will pass only through anti-symmetric 
configurations, that is configurations for which $x_{10}-x_{10-p}=x_p$ and 
$z_{10-p}=-z_p$. This is well verified in Fig. \ref{zdispNodes}.

To find an equilibrium configuration for the chain, we first fix $\bx_5$ 
and as a consequence $\bx_6$. Observe that, due to the symmetry of the 
configuration, this is equivalent to fixing $\theta_5$ and $r_5$ that from now 
on we will call $\theta$ and $r$. Once $\bx_5$ and $\bx_6$ are fixed, we can 
use \eqref{dyna} to compute $\bx_7$ up to $\bx_{10}$ and as a consequence we 
get a full configuration for the chain.

At this point there is no need for $z_{10}$ to be 0. For a range of 
value of $r$ we search for the value of $\theta$ for which $z_{10}=0$. Finally 
we relate back $x_{10}$ to the strain $\delta$. We call this procedure the {\it 
shooting method} to find numerically an equilibrium configuration. Observe that 
high iterations of $f$ in \eqref{dyna} potentially present stability issues. 
This is the reason why we decided to compute only anti-symmetric configurations 
starting from the center of the chain. This is also the reason why the 
application of this methods to longer chains may require a more refined 
analysis of the dynamical system generated by $f$.

\begin{figure}
\centering
\subfigure[]{
\includegraphics[width=0.45\textwidth]{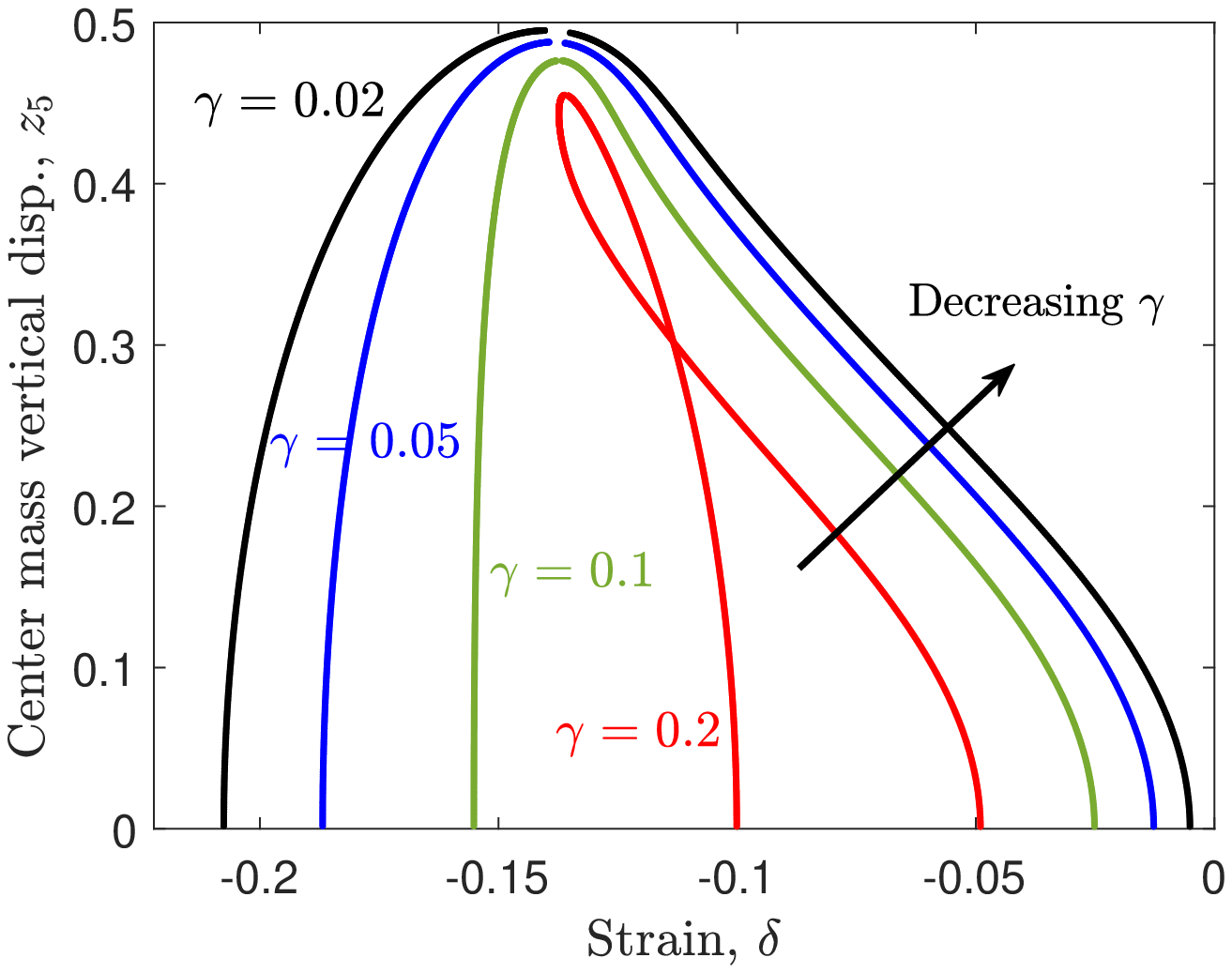}
\label{kg_compareTM}
}
\subfigure[]{
\includegraphics[width=0.45\textwidth]{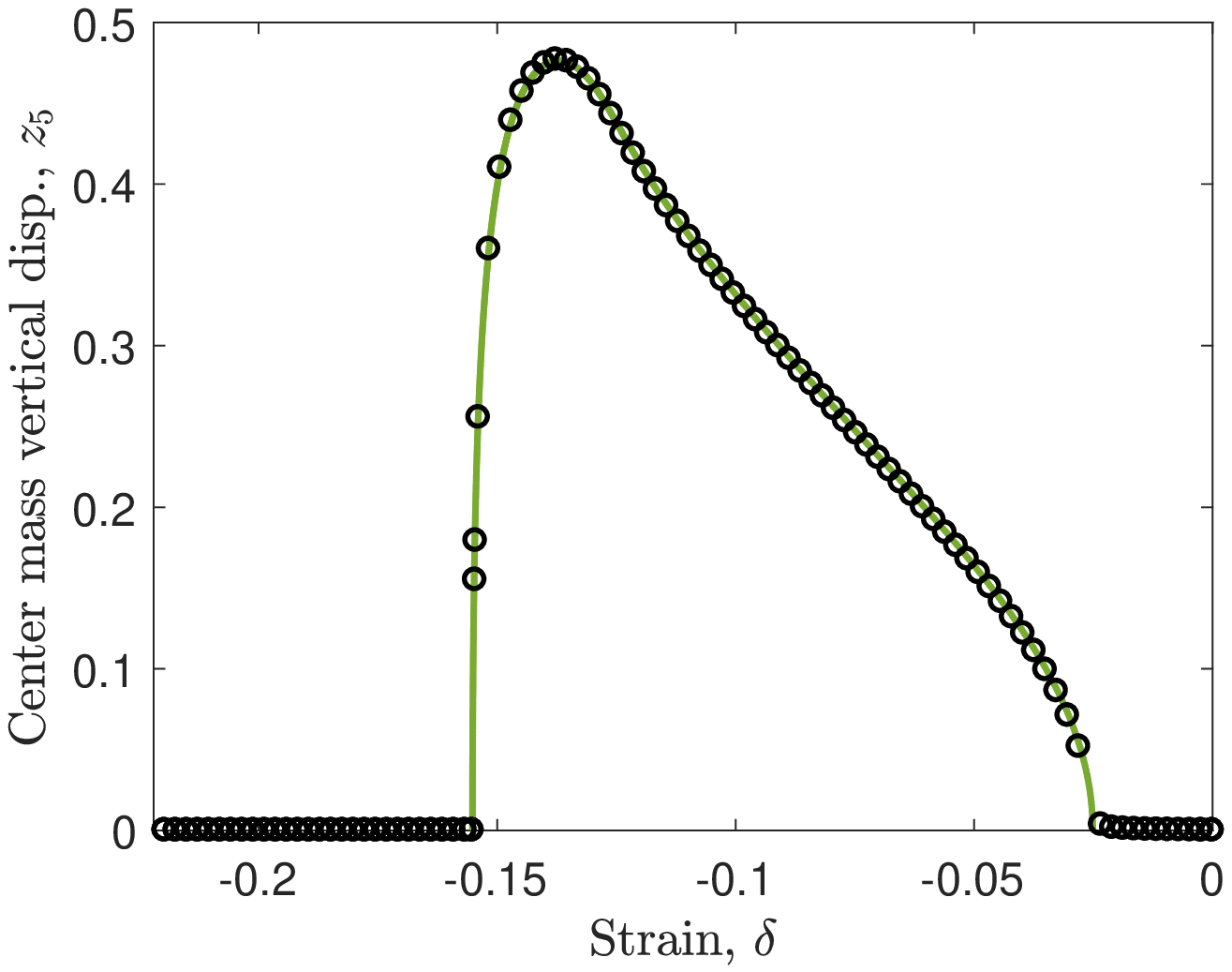}
\label{kg_1e-1_TM}
}
\subfigure[]{
\includegraphics[width=0.45\textwidth]{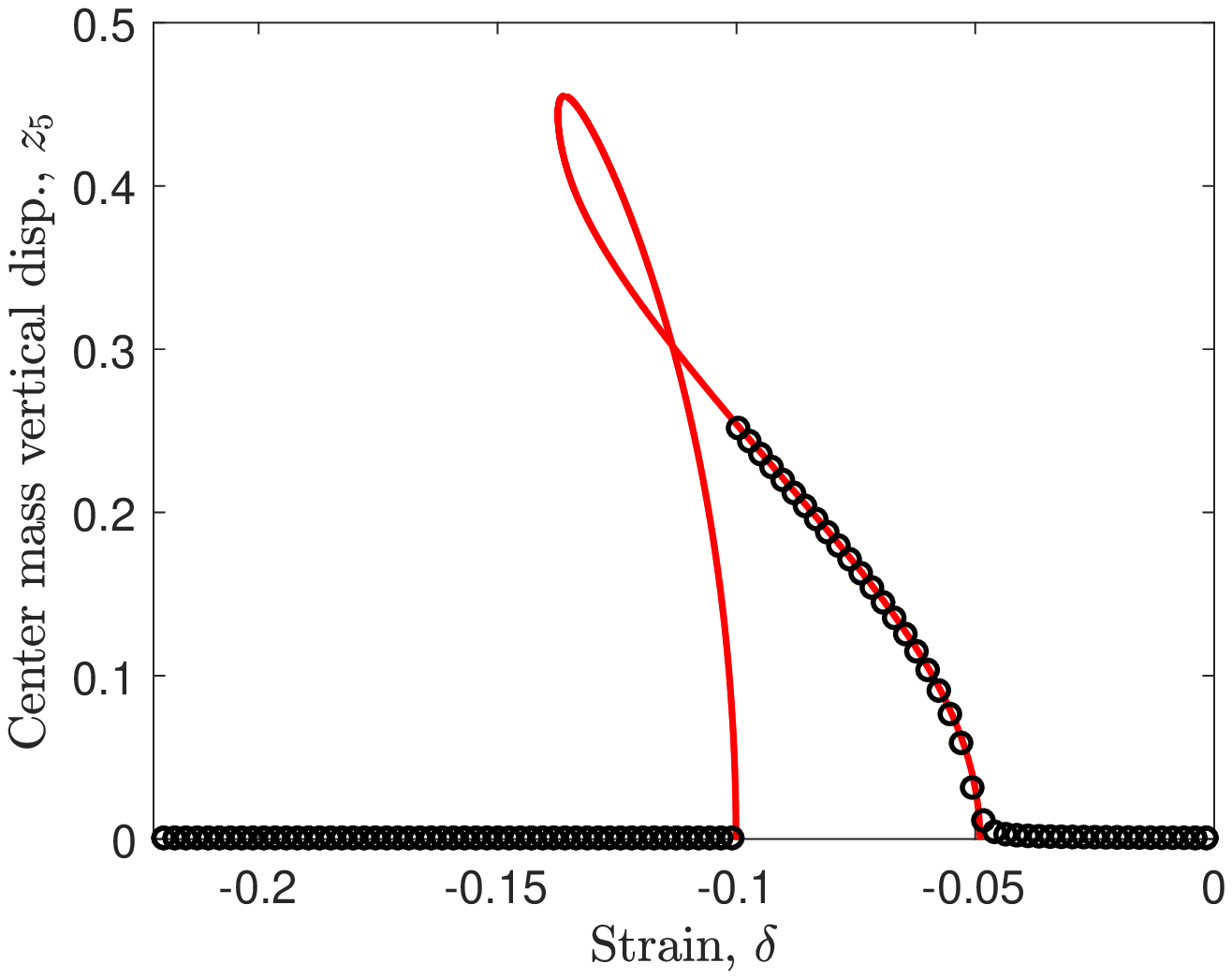}
\label{kg_2e-1_TM}
}
\subfigure[]{
\includegraphics[width=0.45\textwidth]{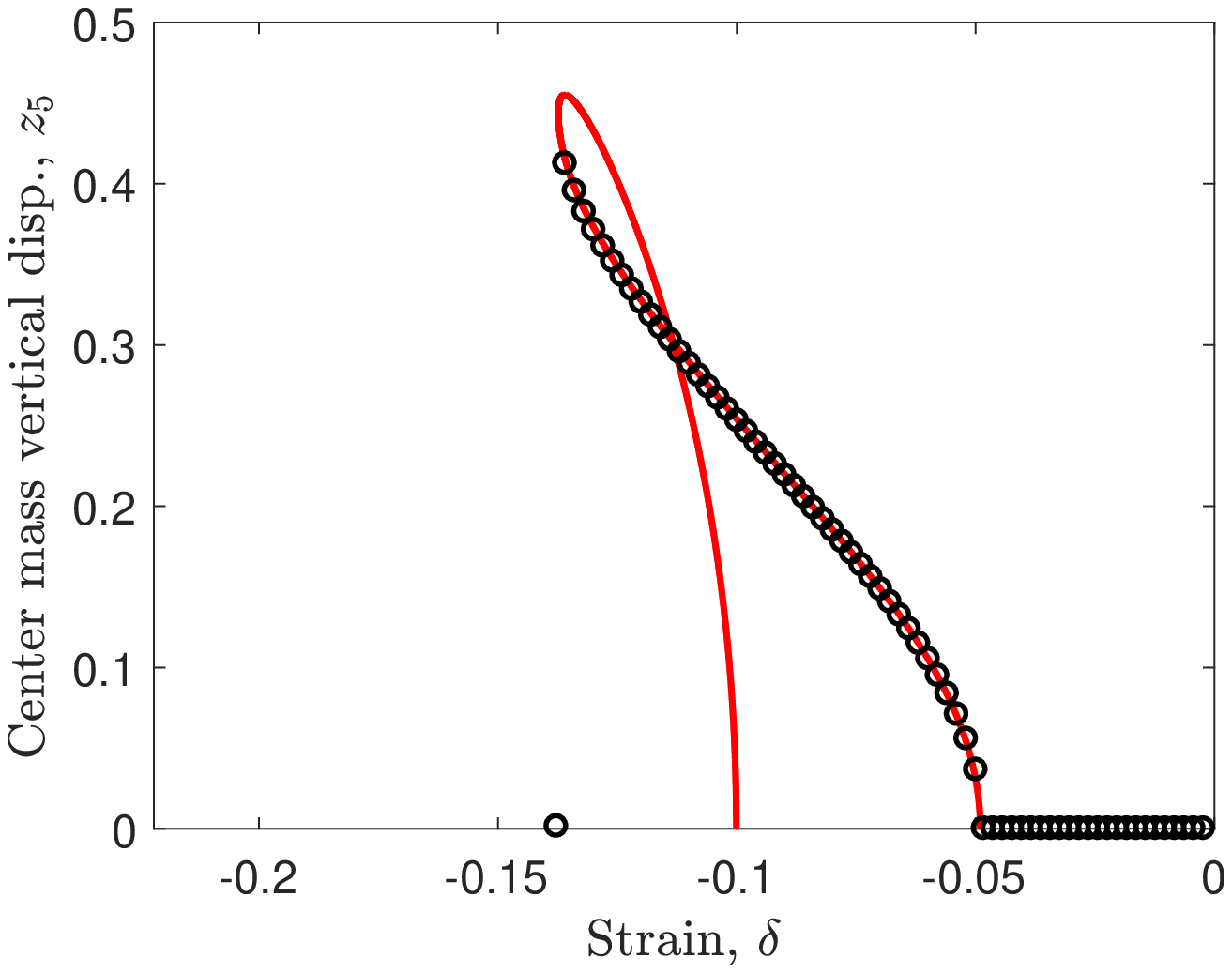}
\label{kg_2e-1_load}
}
\caption{(a) Vertical displacement of the center mass with the axial position of 
the last mass during the folding process for different stiffness parameter 
$\gamma$ values. The Newton-Raphosn method solution (black circles) match the 
shooting method solution (curves) in both loading and unloading for $\gamma = 
0.1$ (a), while the displacement jumps after the onset of instability in both 
loading (c) and unloading (d) for $\gamma = 0.2$.}
\label{Comparekg_tm}
\end{figure}

To gain further insight into  the difference between the high and low stiffness 
parameter response (Fig.~\ref{zdispNodes}) and to understand the displacement 
jump in  Fig.~\ref{zdisp_high}, let us examine the solution as the center spring 
folds. Figure~\ref{kg_compareTM} displays the vertical displacement of the 
center mass $z_5$ with compressive strain $\delta$ for $4$ distinct stiffness 
parameter values. These curves are obtained from the shooting method solution 
described above. For low $\gamma$, there is a single solution for each strain 
value in the range $\delta \in [-2/9,0]$. However, as the stiffness parameter 
$\gamma$ increases, there are three possible solutions in a range of $\delta$ 
values. This multiplicity of solutions is evident as the curve bends backward 
yielding two stable and one unstable solution for each $\delta$ value. By 
comparison, in Fig.~\ref{kg_1e-1_TM} we show with black circles the stable 
solution computed with the previously described Newton-Raphson approach,
as the lattice is stretched from the folded to the unfolded configuration. 
Obviously, there is in excellent agreement with the shooting method  solution 
(solid curve) over the entire range of the folding to unfolding transition. 
Furthermore, the Newton-Raphson method solution of lattice compression 
from the unfolded to the folded configuration also matches with this solution. 

Let us now examine the response of a lattice having a higher stiffness parameter 
($\gamma = 0.2$). Figure~\ref{kg_2e-1_TM} displays the Newton-Raphson 
method solution (black circles) as this lattice is stretched from the folded to 
the unfolded configuration.
As we discussed earlier, the solution is unstable for high $\theta$ values and 
our Newton-Raphson method solution jumps to a stable branch, which 
corresponds to a lower $\theta$ value. As the displacement increases, the 
solution follows this branch until $\theta = 0 $ at around $\delta = -0.049$ 
after which the vertical displacement of all the nodes is zero. Finally, let us 
observe the behavior of this lattice as it is compressed to deform from the 
unfolded to the folded configuration. Figure~\ref{kg_2e-1_load} displays the 
Newton-Raphson method solution for this case. The Newton-Raphson 
method solution matches with the shooting method solution until the onset of 
instability, when the tangent to the red curve in the figure becomes vertical. 
Beyond this point, our Newton-Raphson solver fails to converge to a stable 
equilibrium solution. The black circle at around $\delta=-0.138$ indicates the 
stable solution which a physical system is likely to reach as the prescribed 
compressive strain increases, which corresponds to a folded configuration and 
then follows this horizontal branch (coinciding with the previous 
unfolding case). Thus we observe how the solution can be path dependent for high 
stiffness parameter $\gamma$ values when there are multiple stable solutions. 

\begin{figure}
\centering
\subfigure[]{
\includegraphics[width=0.45\textwidth]{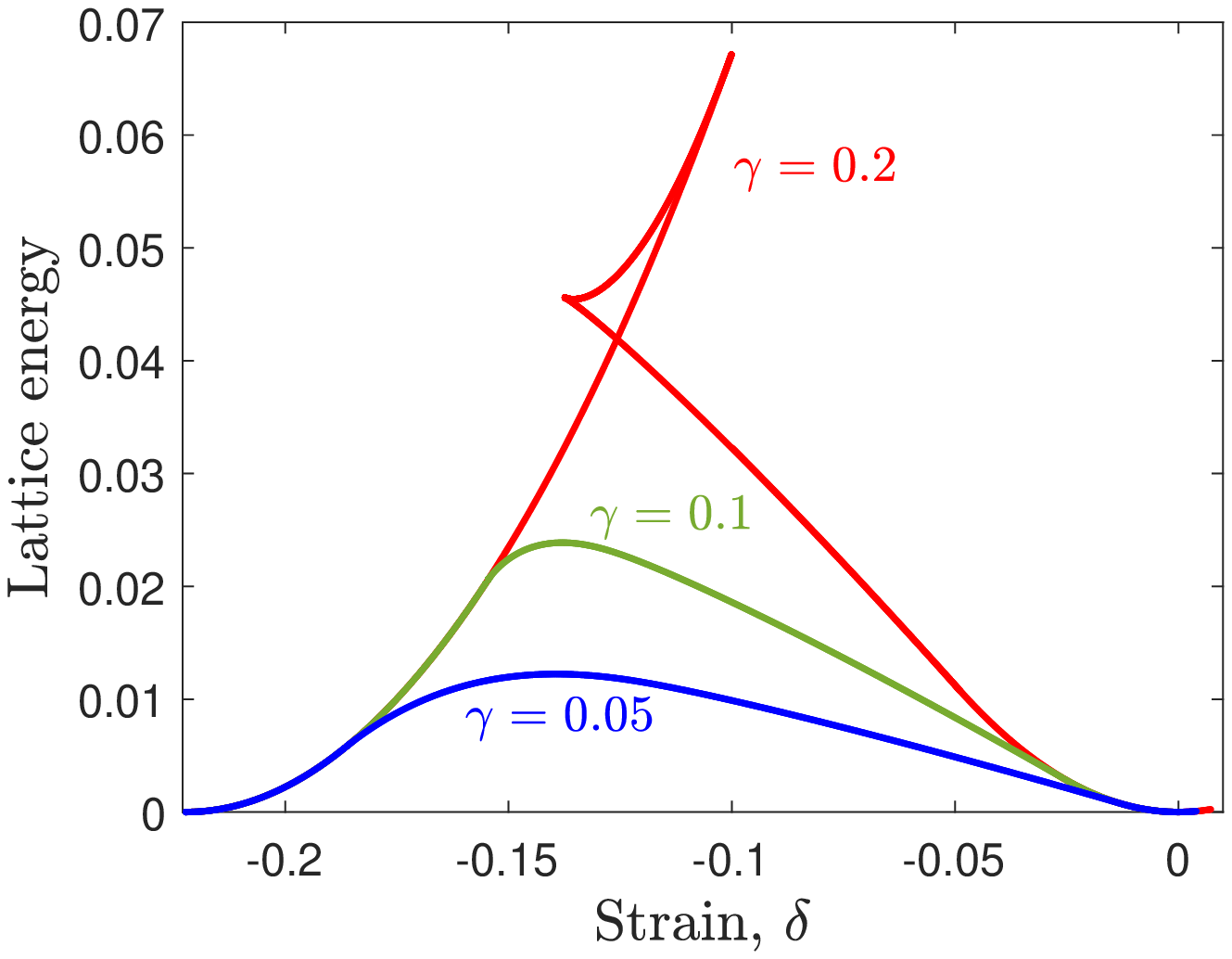}
\label{EnergyLat}
}
\subfigure[]{
\includegraphics[width=0.45\textwidth]{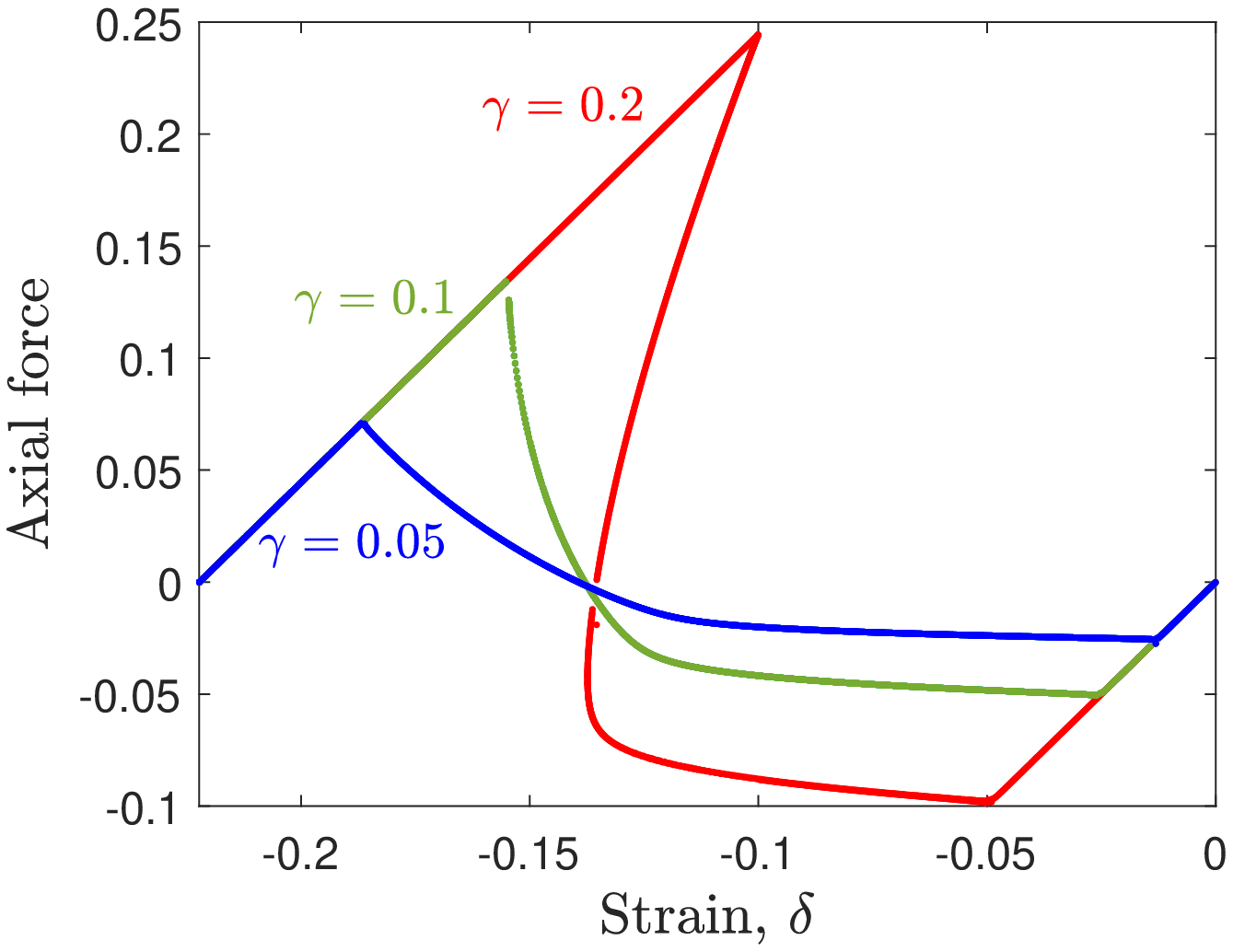}
\label{ForceLat}
}
\caption{(a) Lattice energy and (b) force acting on the ends of the lattice as 
it deforms for different stiffness parameter values. As the stiffness increases, 
the force is not unique in a range of displacements and allows for hysteresis 
during the folding process.}
\label{EnerForce}
\end{figure}

Having provided evidence for the existence of multiple stable solutions, let us 
see how 
the lattice energy and force vary as the lattice deforms from one equilibrium 
configuration to another. Figure~\ref{EnergyLat} displays the lattice energy in 
the vertical axis as it deforms from the folded to the unfolded configuration. 
The strain $\delta$ on the horizontal axis ranges from $-2/9\simeq-0.22$ to 
$0$. 
Three curves corresponding to distinct values of stiffness parameter $\gamma$ 
are shown. In each case, the lattice energy is zero at both the folded and 
unfolded configurations, resulting in a double well potential in this range of 
deformations. At low $\gamma$ values, there is a single energy value for each 
strain $\delta$, while at high $\gamma$, there are multiple energy values 
consistent with the observations above. For $\gamma = 0.2$, we observe that the 
energy curve has three segments and folds back at around $\delta = -0.138$ and 
$-0.1$. The middle segment joining the two end segments corresponds to the 
unstable branch. 

Let us now illustrate how the corresponding horizontal  force $F$ acting on the 
chain varies during the folding process. It is defined as the derivative 
of the potential energy with respect to the chain length, given by
\begin{equation*}
F = \dfrac{1}{N-1} \dfrac{\partial E}{\partial \delta}
\end{equation*}
and it is computed by taking the numerical derivative of the energy illustrated 
in Fig.~\ref{EnergyLat}. Figure~\ref{ForceLat} illustrates the force for the 
same three stiffness parameter values $\gamma$. In all cases, the force is zero 
in the folded and unfolded configurations at $\delta = -2/9$ and $\delta = 0$. 
For low stiffness parameter values ($\gamma = 0.05 $ and $\gamma=0.1$ in the 
figure), there is again a unique force for each strain value $\delta$. The force 
displacement response has three segments: two linear and one segment in the 
middle joining them. The linear segments correspond to affine horizontal 
deformation about the folded and unfolded configurations. The middle segment 
corresponds to the rotation of the center spring as $\theta$ varies from $0$ to 
$\pi$ and the effective stiffness of the chain is negative in this regime as the 
force decreases with increasing prescribed displacement or tensile strain. 
Furthermore, there are multiple values of force in a range of strains $\delta$ 
for the $\gamma=0.2$ high stiffness parameter case. In this range of 
displacements, there are two stable solutions and one unstable branch. The 
force-displacement response illustrates the potential for achieving hysteresis 
in our lattice in two ways. By prescribing displacement and alternating between 
the folded and unfolded configurations in a lattice with high stiffness 
parameter $\gamma$, the lattice follows a different path around the unstable 
branch. For example, in Fig.~\ref{ForceLat}, the force jumps down from $0.24$ to 
$-0.088$ while unfolding and it jumps up from $-0.042$ to $0.17$ while folding. 

\begin{rem}
We finally briefly remark on the alternative possibility of inducing hysteresis 
by exploiting the negative stiffness zone. Instead of prescribing displacement, 
if we prescribe a horizontal force at the ends of the chain, then the lattice 
solution will jump at the end of the first segment for all the three stiffness 
values, for example at a force level $F = -0.026$ from $\delta = -0.013$ while 
folding and at a force level $F = 0.0710$ from $\delta = -0.187$ while 
unfolding. Thus our results illustrate the potential for hysteresis and 
localization guided  programmable smart materials in lattice based media. 
\end{rem}

\subsubsection{Arc length method for four mass chain}\label{4masschain}

Here we consider the simplest case of $N=4$, that is a chain of $4$ point 
masses 
anchored at its extremes.  
As in previous sections, the unfolded and folded configurations of the four 
masses are
$$(0,0),\,\ (1,0),\,\ (2,0),\,\ (3,0)\ \quad \text{and} \quad
(0,0),\,\ (1,0),\,\ (0,0),\,\ (1,0)\ .
$$
We can (and will) parametrize the $x$-coordinate of the last mass as $3l$, with 
$1/3\le l\le 1$,
ranging through the folded and unfolded states.  Alternatively, we can use the 
strain, which
is related to $l$ by the linear relation 
$$\delta=l-1\in [-2/3, \ 0] .$$

Calling $\bx_i=(x_i,z_i)$ the position of the $i$-th 
point mass, the energy of the system, expressed in nondimensional form by 
normalizing it by a reference energy $k_0 a^2/2$ is
\begin{equation}\label{Ham}
E = \frac{\gamma}{2}\sum_{i=1}^4 
z_i^2\,+\,\frac{1}2\sum_{i=1}^{3} (\|\bx_{i+1}-\bx_{i}\|-1)^2
\end{equation}
where the first sum is the potential 
energy of the vertical restoring force and the last sum is the potential energy 
of 
the springs. In this last sum $\bx_1$ and $\bx_{4}$ represent the fixed 
extreme of the chain, namely 
$\bx_1=(0,0)$, $\bx_4=(3l,0)$.

Now, with $N=4$, we assume that the equilibrium 
position satisfies $x_3=3l-x_2$ and $z_3=-z_2$.
Let us call $r$ the length of the segment from the middle
point to the location of the second point mass $\bx_2$, and call $\theta$
the angle formed between the middle point and the horizontal.
This way, the 4 point masses get located at the points:
$$\bx_1=(0,0),\,\ \bx_2=(3l/2-r\cos \theta,\ r\sin \theta) ,\,\ 
\bx_3=(3l/2+r\cos \theta,\ -r\sin \theta),\,\ \bx_4=(3l,0)\ .$$

Clearly we have only the variables $r$ and $\theta$ to consider and the 
potential to minimize becomes
\[
E(r,\theta) = \gamma 
r^2\sin^2\theta+\left(\sqrt{r^2-3lr\cos\theta+\frac{9l^2}4}-1
\right)^2+\frac12\left(2r-1\right)^2
\]
so that by taking derivatives we get the necessary conditions
\begin{equation}\begin{split}\label{partials}
E_r\equiv 2 \gamma r\sin^2\theta+\left(2r-3l\cos\theta\right)\left(1-\left( 
r^2-3lr\cos\theta+\frac{9l^2}4\right)^{-\frac12}\right)+2(2r-1)=&0 \\
E_{\theta}\equiv 2 \gamma r^2\sin\theta\cos\theta+3lr\sin\theta\left(1-\left( 
r^2-3lr\cos\theta+\frac{9l^2}4\right)^{-\frac12}\right)=&0\ .
\end{split}\end{equation}
Now, observe that there are two solutions valid for all $\gamma$: 
\begin{description}
\item[Unfolded] $r=l/2$ and $\theta=0$;
\item[Folded] $r=\frac{2}{3}-\frac{l}2$ and $\theta=\pi$.
\end{description}

Unlike our previous numerical experiments, here we implement a classic 
pseudo-arc-length continuation algorithm starting with the trivial solution 
branch $(l/2,0)$ for $l$ ranging from $l=1$ to $l=1/3$, and monitoring branch 
points\footnote{Namely, values where there is another solution branch passing 
through them.} originating from this branch.  Further following this bifurcating 
branch, we will eventually connect the branch $(l/2,0)$ to the other solution 
branch $(\frac{2}{3}-\frac{l}2,\pi)$.  Stability monitoring is done by looking 
at the Hessian eigenvalues.  In the graphs below, we show what happens for 
several different values of $\gamma$.  The most insightful visualizations are 
obtained displaying the values of $\cos(\theta)$, with the two trivial branches 
corresponding to the values $1$ and $-1$; when not confusing, we also display 
the height value, as in the figures of the previous sections. We will show in 
bold-face the stable portions of the equilibria branches.  In spite of different 
values of $\gamma$, the picture we obtain in this simple case of $N=4$ is quite 
similar to what we observed in the previous sections for $N=10$.

\medskip

\begin{figure}
\centering
\subfigure[]{
\includegraphics[width=0.45 \linewidth]{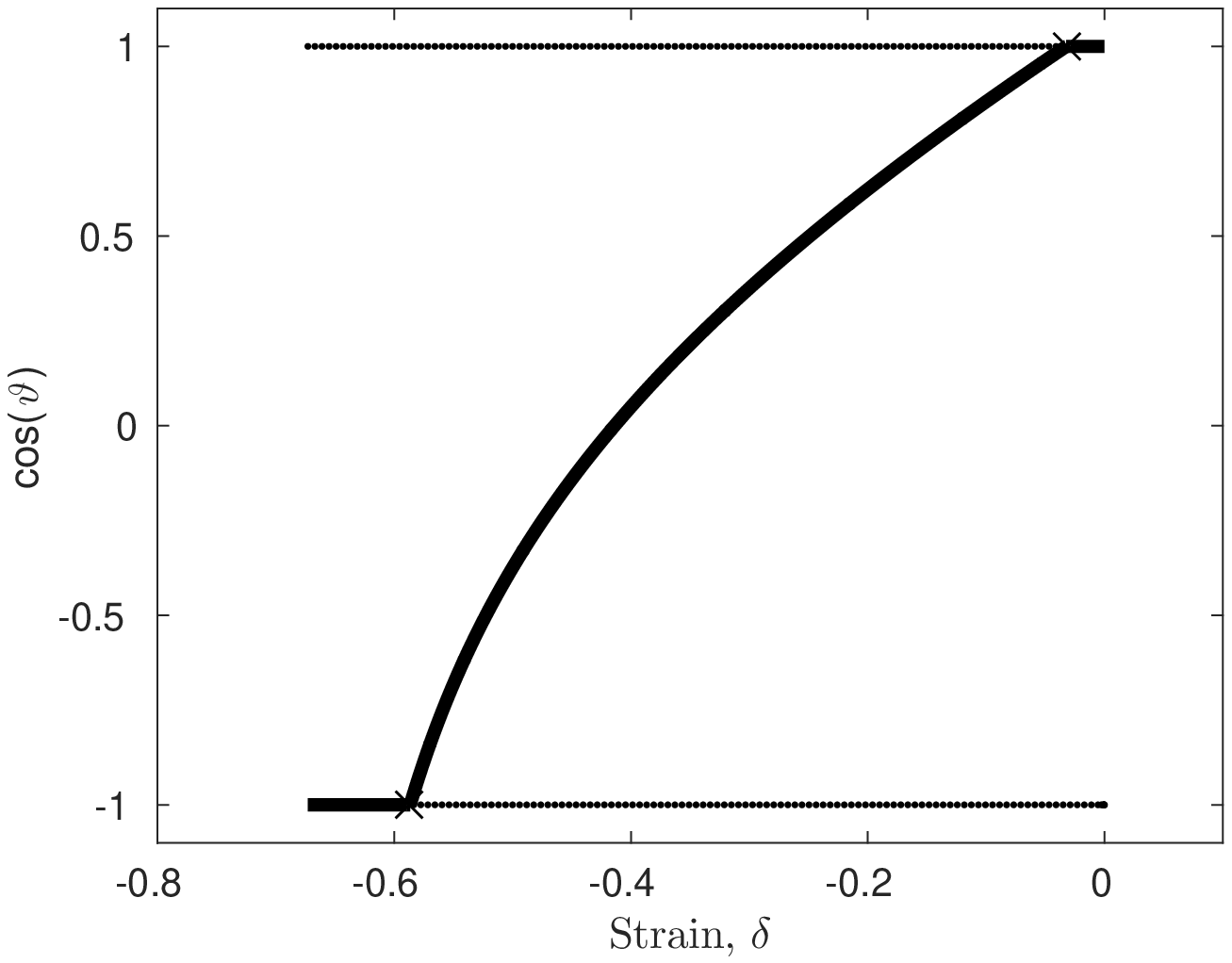}
\label{Figsfig1}
}
\subfigure[]{
\includegraphics[width=0.45 \linewidth]{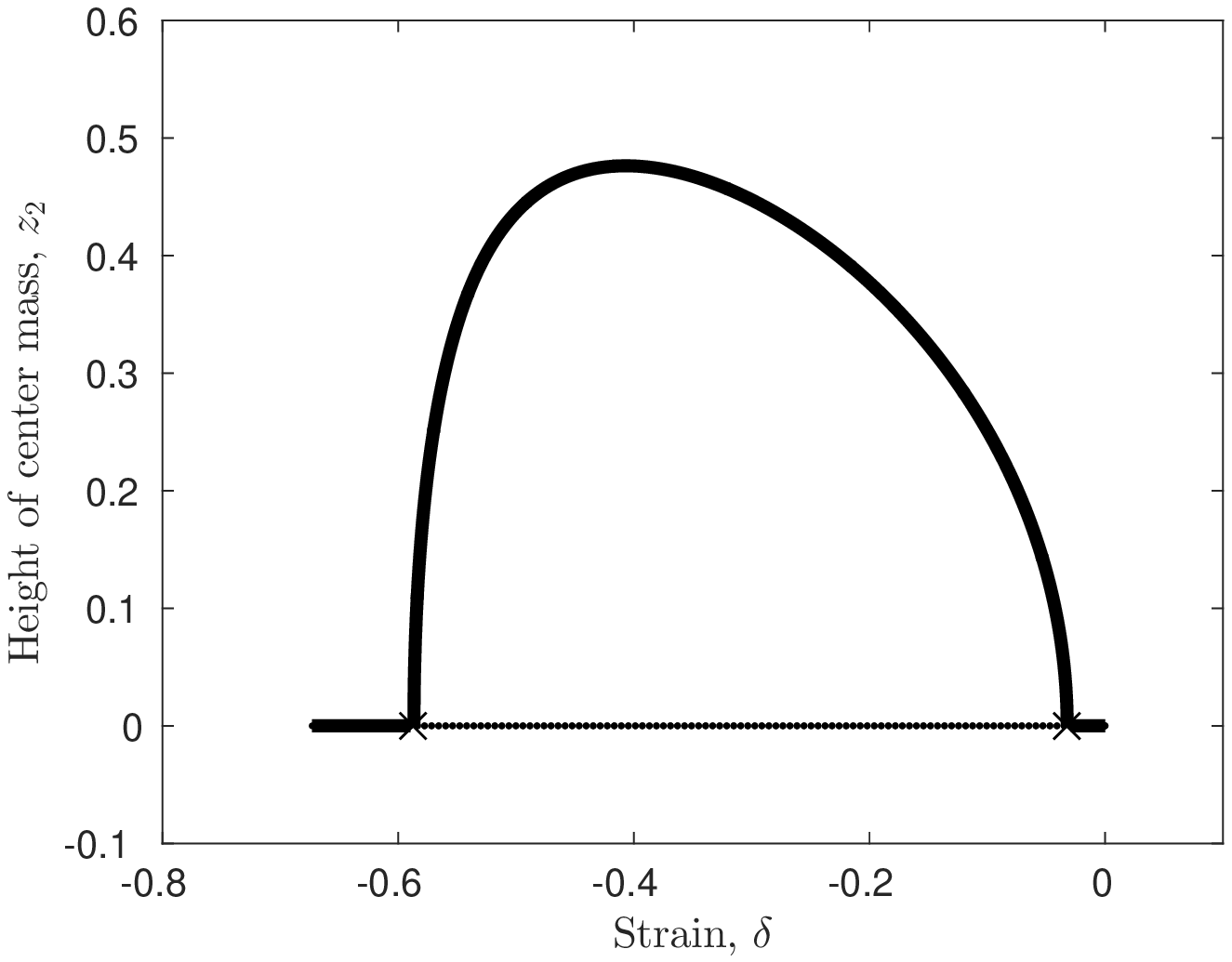}
}
\caption{$\gamma=0.1$.  Stable equilibria connection between unfolded and 
folded 
configurations.}
\end{figure}

For $\gamma$ small ($\gamma=0.1$ in Figure \ref{Figsfig1}-(b)), we have a stable 
connection of equilibria configurations between the two trivial branches of 
equilibria. That is, for each value of the strain in $[-2/3,0]$, there is always 
only one stable stationary solution.

\begin{figure}
\centering
\subfigure[]{
\includegraphics[width=0.45 \linewidth]{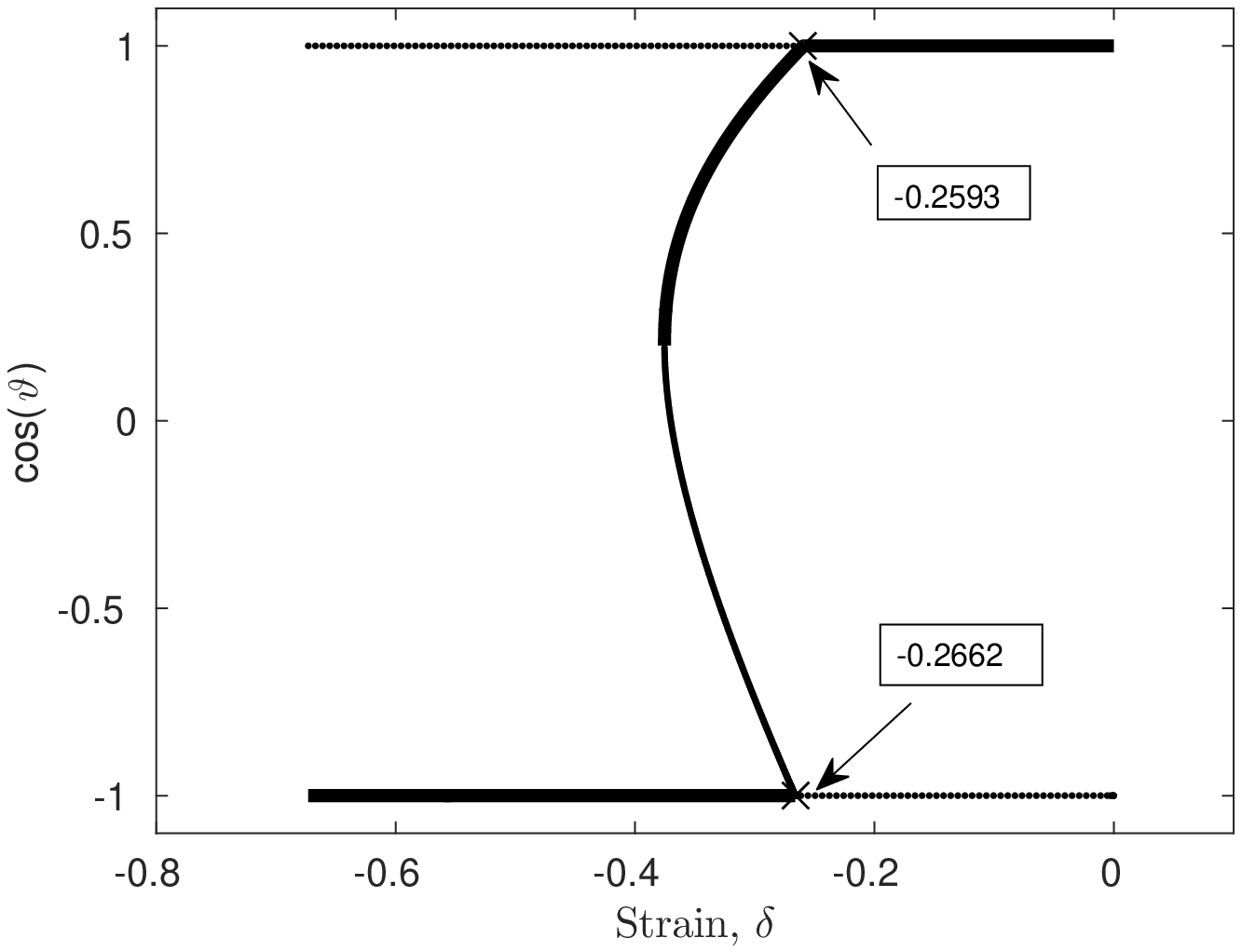}
\label{Figsfig2}
}
\subfigure[]{
\includegraphics[width=0.45 \linewidth]{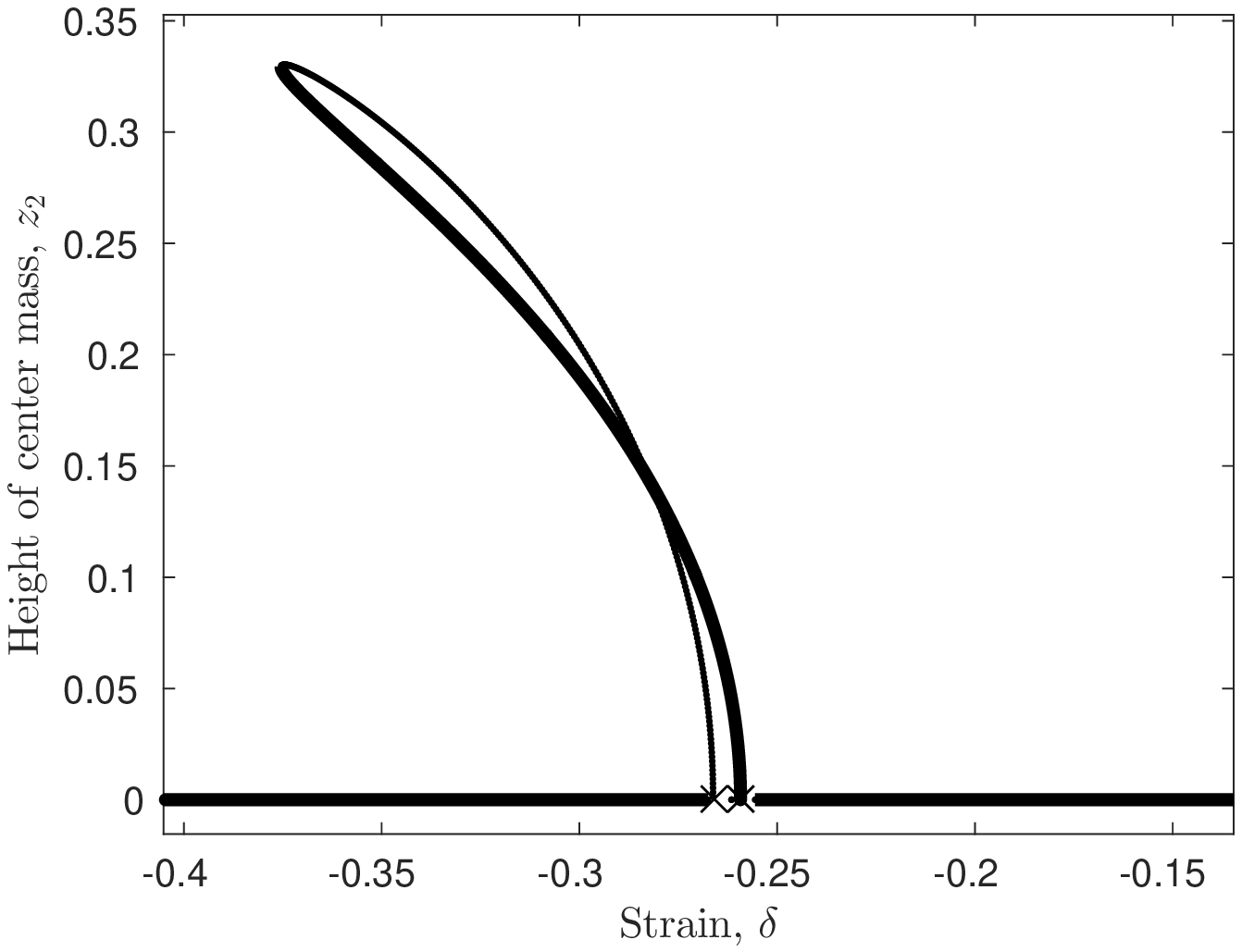}
}
\caption{Partially stable equilibria connection between unfolded and folded 
configurations.
Hysteretic behavior.}
\end{figure}

As $\gamma$ grows, the stable connection between the two trivial branches 
develop a fold and there are two stable equilibrium solutions for a range of 
values of the strain: one of the trivial branches and (portion of) the 
connecting branch; see the case of $\gamma=1.05$ in Figure \ref{Figsfig2}-(b).  

\begin{figure}
\centering
\subfigure[]{
\includegraphics[width=0.45 \linewidth]{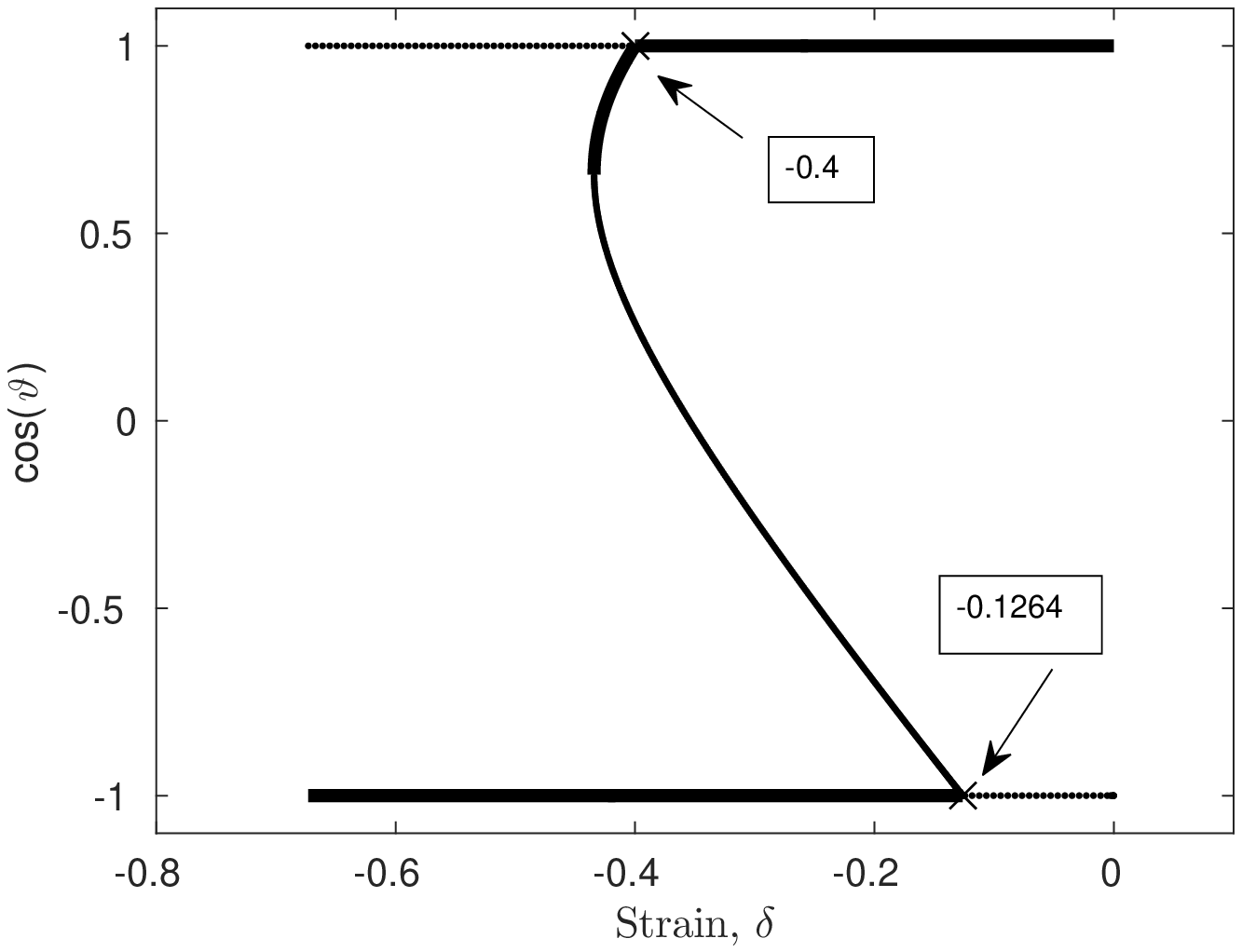}
\label{Figsfig3a}
}
\subfigure[]{
\includegraphics[width=0.45 \linewidth]{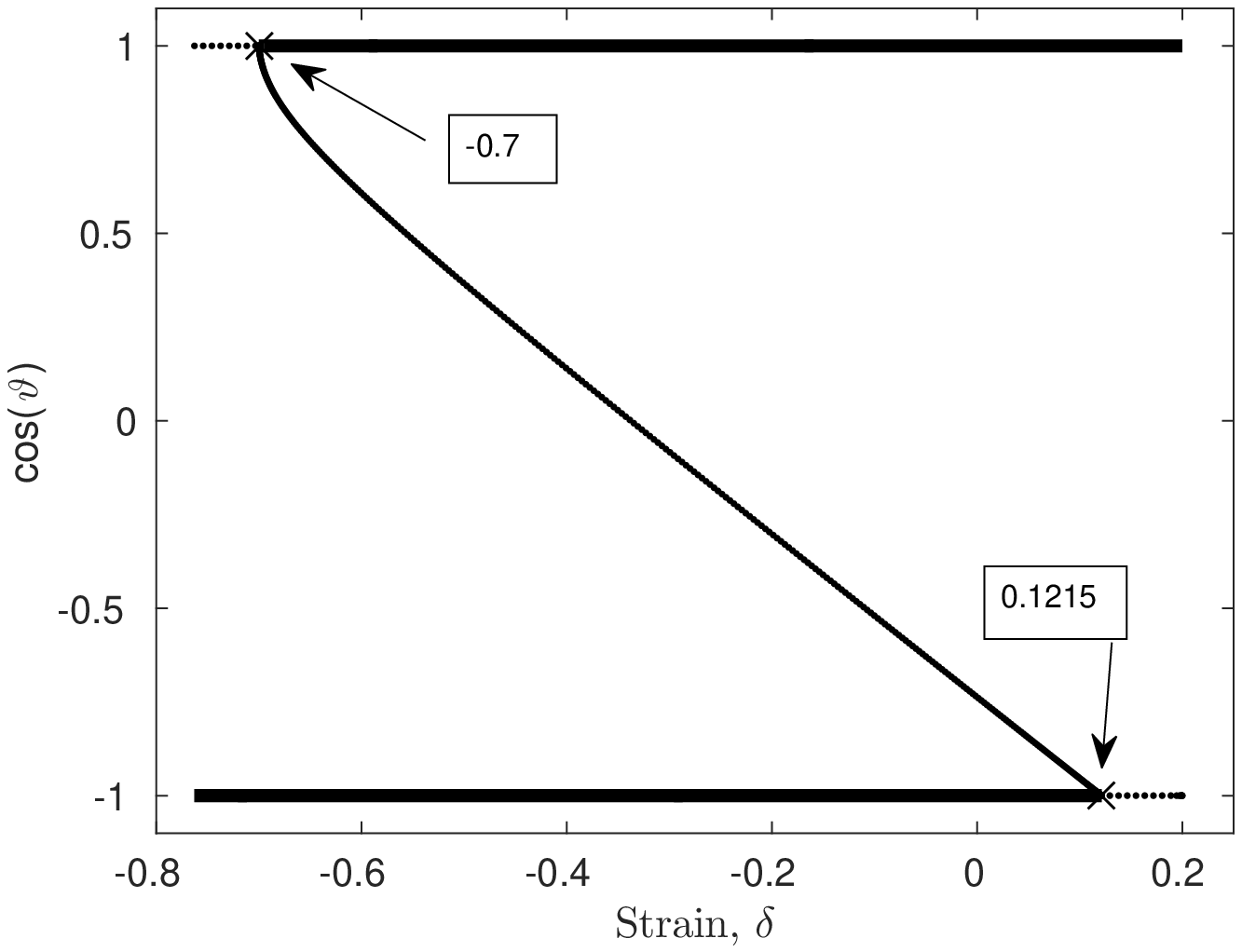}
\label{Figsfig3b}
}
\caption{(a) $\gamma=2$. Both folded and unfolded branches are stable for a 
range of strain values. (b) $\gamma=7$. As $\gamma$ becomes large, the branch of 
equilibria connecting folded and unfolded states is fully unstable; the states 
themselves are stable for all strain values.}
\end{figure}

At the same time, as we increase $\gamma$ further, there is a range of values of 
the strain where both trivial branches are stable, as well as part of the 
connecting branch of equilibria; see $\gamma=2$ in Figure \ref{Figsfig3a}.  
Eventually, for $\gamma$ sufficiently large ($\gamma=7$ in Figure 
\ref{Figsfig3b}), the connecting branch is made up entirely of unstable 
equilibria and the branch points on the trivial branches occur outside the range 
of values of the unfolded/folded configurations. 

\begin{rem}
As an alternative to the continuation technique we used, we also
double checked our results by a more algebraic technique, obtaining the same 
results.
Dividing the second equation in \eqref{partials} by $r\sin\theta$, and 
with some algebra, \eqref{partials} rewrite as
\begin{equation}\begin{split}\label{sol2}
(2 \gamma r\cos\theta+3l)^2\left(r^2-3lr\cos\theta+\frac{9l^2}4\right)=&9l^2 \\
(2r-3l\cos\theta)\gamma r\cos\theta-3l( \gamma r\sin^2\theta+2r-1)=&0\ .
\end{split}\end{equation}
From \eqref{sol2}, the second equation becomes
\[
 2 \gamma r^2\cos\theta-3l(( \gamma+2)r-1)=0
\]
that is
\begin{equation}\label{cos}
 \cos\theta=\frac{3l((\gamma+2)r-1)}{2\gamma r^2}
\end{equation}
that replaced in the first of \eqref{sol2} gives
\[ 
\left(3l((\gamma+2)r-1)+3lr\right)^2\left(8\gamma r^3-36l^2((\gamma+2)r-1)
+18 \gamma l^2r\right)-72 \gamma l^2r^3=0
\]
or
\begin{equation*}
18 l^2\, -\, 9 l^2(5\gamma+16)r\, +\, 18 l^2(\gamma+3)(2\gamma+7)r^2\, -\, 
9l^2(\gamma+3)^2(\gamma+4)r^3\, -\, 8\gamma(\gamma+3)r^4\, +\, 
4\gamma(\gamma+3)^2r^5  =0\ .
\end{equation*}
From the roots of this quintic, which we do by finding them as eigenvalues of 
the
associated companion matrix (and keeping only the real ones 
for which the relative $\cos \theta$ given by \eqref{cos} is in $[-1,1]$), we 
obtain
the possible equilibrium solutions.
\end{rem}

\section{Conclusions}\label{concSec}

To understand, predict and control localization patterns in lattices, we 
considered a square lattice on an elastic substrate. We exemplified how the type 
of instability changes from in-plane to out of plane with increasing stiffness 
parameter. The latter instability evolves to a localized deformation with 
increasing strain and it is investigated in detail using a one-dimensional 
lattice. Using a shooting method based approach, we identified the stable 
branches of deformation as the lattice deforms from one stable configuration to 
another with a localized deformation. For lattices with low stiffness parameter, 
there is only one stable solution as the localized pattern forms, while if this 
spring stiffness exceeds a critical value, there are two stable solutions in a 
range of displacements. The lattice energy and force response demonstrate the 
potential for inducing hysteresis due to the existence of the two stable 
solutions. Finally, we considered the simplest model possible, $4$ point masses 
with the two endpoints fixed. In this case, we used a classical continuation 
technique and were able to confirm the transition from unfolded to folded 
quasi-static equilibrium configurations. 

In summary, we have illustrated that, in contrast to many other existing works 
where localization is unpredictable and sensitive to the presence of defects, 
localized deformation can be predicted precisely based on geometry and symmetry 
considerations. Our analysis framework may open avenues for controlled design of 
interfaces for waveguiding and programmable smart materials applications. 

\bibliographystyle{unsrt}
\bibliography{paper}

\end{document}